\documentclass{article}
\usepackage[T5]{fontenc} 
\usepackage[fontsize=12pt]{scrextend}
\usepackage[a4paper, right=1.5cm, left=2cm, top=2cm, bottom=2cm]{geometry} 
\usepackage{graphics} 
\usepackage{float} 
\usepackage{tikz} 
\usetikzlibrary{calc} 
\usepackage{colortbl}
\usepackage{setspace}
\usepackage{hhline}
\usepackage{soul,color}
\usepackage{comment}
\usepackage{hyperref}
\usepackage{indentfirst}
\usepackage{titlesec}
\usepackage{afterpage}
\usepackage{caption}
\usepackage{titletoc}
\usepackage{mathtools}
\usepackage{booktabs}
\usepackage{multirow}
\usepackage{xcolor}
\usepackage[utf8]{inputenc}
\usepackage{amsmath,amssymb,amsfonts}
\usepackage{tabularx}
\usepackage{algorithm}
\usepackage{algpseudocode}
\usepackage{booktabs}
\usepackage{graphicx}
\usepackage{textcomp}
\usepackage{xcolor}
\usepackage{rotating}
\usepackage{adjustbox}
\usepackage{subfig}
\usepackage[font=small,labelfont=bf]{caption}
\DeclareUnicodeCharacter{0301}{\ignorespaces}
\usepackage{cite}
\begin{document}


\begin{center}
    \textbf{
    LiteNeXt: A Novel Lightweight ConvMixer-based Model with Self-embedding Representation Parallel for Medical Image Segmentation
    }\\
    \medskip
    \textit{Ngoc-Du Tran, Thi-Thao Tran$^{*}$, Quang-Huy Nguyen, Manh-Hung Vu, Van-Truong Pham}\\
    $^{*}$Corresponding author:  thao.tranthi@hust.edu.vn
\end{center}
\textit{Department of Automation Engineering, School of Electrical and Electronic Engineering, Hanoi University of Science and Technology}\\
\medskip

\section*{Abstract}

The emergence of deep learning techniques has advanced the image segmentation task, especially for medical images. Many neural network models have been introduced in the last decade bringing the automated segmentation accuracy close to manual segmentation. However, cutting-edge models like Transformer-based architectures rely on large scale annotated training data, and are generally designed with densely consecutive layers in the encoder, decoder, and skip connections resulting in large number of parameters. Additionally, for better performance, they often be  pretrained on a larger data, thus requiring large memory size and increasing resource expenses.  
In this study, we propose a new lightweight but efficient model, namely LiteNeXt, based on convolutions and mixing modules with simplified decoder, for medical image segmentation. The model is trained from scratch with small amount of parameters (0.71M) and Giga Floating Point Operations Per Second (0.42). To handle boundary fuzzy as well as occlusion or clutter in objects especially in medical image regions, we propose the Marginal Weight Loss that can help effectively determine the marginal boundary between object and background. Additionally, the Self-embedding Representation Parallel technique is proposed as an innovative data augmentation strategy that utilizes the network architecture itself for self-learning augmentation, enhancing feature extraction robustness without external data. Experiments on  public datasets including Data Science Bowls, GlaS, ISIC2018, PH2, Sunnybrook, and Lung X-ray data show promising results compared to other state-of-the-art CNN-based and Transformer-based architectures.  Our code is released at: \href{https://github.com/tranngocduvnvp/LiteNeXt---A-Lightweight-Mixer-based-Model}{https://github.com/tranngocduvnvp/LiteNeXt}.

\textbf{Keywords.} Biomedical Image Segmentation, Mixer Models, Self-embedding Representation Parallel, Marginal Weight Loss, LiteNeXt.

\newpage 

\section{Introduction}
In medical imaging, segmentation is a crucial and often performed activity that allows for the extraction of precise structural information about the areas of interest. Manual segmentation is tedious, time-consuming, and requires experienced experts and doctors to give accurate results, so applying artificial intelligence to build the automatic segmentation framework in medical diagnosis is an urgent task. Deep learning models have outperformed conventional techniques in recent years on a variety of computer vision tasks, including object identification, picture segmentation, and image classification \cite{unet,fcn_long, pattern_fewshot}. The application of automation in the learning process to medical imagine identification has grown in favor. In the field of medical image analysis, segmentation models could help shorten the time to determine the damaged areas and interested tissues from images like brain tumor \cite{dong2017automatic,deep_cascade_brain,isensee2018brain}, brain \cite{pattern_semi}, left ventricles from cardiac magnetic resonance images \cite{blunet,acdc},   dermoscopic skin lesions,\cite{nham22,skin_diagnosis}, cell microscopy images \cite{nuclei_evaluation,nucleus_exemplars}, otoscope ear-drum images  \cite{earnet}, whole heart \cite{pattern_semi}. Thus minimizing human subjective errors during object delineation, and also helps doctors make accurate diagnoses as well as develop effective treatment regimens for patients.

In typical deep learning segmentation models, an input image is passed through an encoder to extract semantic information as feature maps. These features will then be decoded through the decoder to create a prediction mask. However, when encoding images with an encoder, although the semantic information of the image is increased, the spatial information of the image is degraded. Therefore, when performing the decoding process, it often does not produce highly accurate results. To solve that problem, Ushape models \cite{unet, attentionunet} have been researched with a design that includes two symmetric U-shaped encoder and decoder branches and shortcut connections between the two branches to solve the problem of spatial information loss during decoding. Ushape models have shown superior performance over traditional FCN \cite{fcn_long} models that do not use symmetric encoder decoder branches like FCN32 \cite{fcn_long}. Although they provide high accuracy, Ushape models have a large number of parameters and high computational costs. 

Designing lightweight models for image segmentation has garnered more attention recently. Lightweight models with fewer parameters and less inference time could be applicable for mobile devices and real applications. One of early lightweight models can be considered is Medical Transformer (MedT) \cite{valanarasu2021medical}, which is first introduced in 2021. This model is based on transformer network architecture with a desire that the model can learn long-range dependencies and highly expressive representations. This model proposes a gated axial-attention model by giving the self-attention module an extra control mechanism. By applying axial attention along the height and width axes, the original self-attention mechanism is effectively modeled with significantly improved computational efficiency. Though impressive with 1.4M parameters, the MedT has a large number of Giga Floating Point Operations Per Second (GFLOPs) that suffer from slow inference time. UNeXt \cite{valanarasu2022unext} is another lightweight model for image segmentation with small GFLOPs and only 1.47M parameters, based on Convolution and multilayer perceptron. The architecture of the model still follows encoder-decoder architecture of UNet \cite{ronneberger2015u} with skip connection but with the other design in each block. In a recent work,  U-Lite \cite{dinh20231m} is introduced with even fewer parameters than MedT and UNeXt models, only 878K parameters. This architecture also follows the symmetric architecture of UNet with encoder, decoder and skip connection but is designed in order to strengthen the power of CNNs while reducing remarkable the number of parameters. The authors proposed an Axial Depthwise Convolution module, which is inspired by the design of Vision Permutator and ConvNeXt. This module uses axial convolution 7x7 that offers a large receptive field and the pointwise convolution to help encode features along the depth of the feature map and change the number of feature map channels flexibly, leading to fewer parameters than traditional convolution.


{In general, lightweight models demonstrate acceptable performance compared to models with dozens of times more parameters on specific tasks. However, most of these models rely on U-shape paths and complex non-parametric connections, requiring large memory to store intermediate tensors for decoding. This hinders memory reuse on devices with limited resources. Additionally, many lightweight designs sacrifice segmentation accuracy due to simplified convolution or attention mechanisms. In this context, we propose LiteNeXt, a lightweight segmentation model that incorporates Self-embedding Representation Parallel (SeRP) as a novel augmentation strategy leveraging network architecture for robustness and Marginal Weight Loss to enhance boundary delineation. These innovations address the limitations of existing lightweight models while ensuring efficient performance.}

Inspired from the advantages of lightweight models, in the current study, we aim to develop a model with a small number of parameters and low computational cost, while achieving accuracy competitive with more complex models such as UNet\cite{ronneberger2015u} {and Transformer-based models. We observe that Transformer-based models like TransUnet}\cite{transunet} {and SwinUnet}\cite{swinunet} {exhibit simple designs yet achieve competitive segmentation performance compared to CNN-based models. A key factor contributing to this difference is the robustness of {their} backbone encoder, which leverages {pretraining} on large datasets. From this observation, we derive two conclusions:} {($i$)The robustness of a model heavily relies on its feature extraction capability; and ($ii$) if a backbone can extract information effectively, the decoder can be simplified to reduce parameters and computational cost without compromising performance. In other words,} if we could design a good enough feature extractor, we could eliminate the multi-layer decoder branch. {Accordingly,} the model's computational cost and parameter quantity may be greatly minimized. {Based on these conclusions, we propose the LiteNeXt model, which features a simplified decoder and the Self-embedding Representation Parallel training strategy (SeRP)} adapted from the BYOL self-supervised learning \cite{byol} {to enhance the robustness of the encoder. Besides, to handle boundary segmentation in medical images, we introduce a marginal weight loss with a reasonable weighting strategy for each object region that helps emphasize on the marginal boundary in the loss function.} 
Details about the sections will be presented in the following sections.

In summary, the main findings are listed as follows:
\begin{itemize}
    \item We propose a novel model called LiteNeXt with only 0.72M parameters but achieves superior performance compared to many other lightweight methods.
    \item A new weighting method in the loss function is introduced called Marginal Weight Loss, which is able to well separate the boundaries between objects in medical images.
    \item A self-supervised training method called Self-embedding Representation Parallel was applied for the first time in the fully-supervised segmentation problem to improve the robustness of the model.    
     Experiments performed on {various} popular medical datasets, which are Data Science Bowl 2018 dataset, GlaS Segmentation dataset, ISIC2018 Lesion Segmentation dataset, PH2 dataset, Sunnybrook cardiac dataset, and Lung X-ray data,  have shown the superior performance of the proposed model and loss function compared to other methods.
    
\end{itemize}

\section{Related Work}

\subsection{Mixer Models}

After the success of Vision Transformers (ViT) \cite{dosovitskiy2020image} on {the} image classification task, a series of studies \cite{pvit, lai2022axial} with the general design of blocks consisting of two parts: spatial mixing and channel mixing along with skip connections have shown impressive results on many vision tasks. Tolstikhin et al. \cite{mlpmixer} first proposed the MLP-mixer model that is capable of mixing both spatial and channel-wise information by using only MLP blocks. Trockman \cite{convmixer} et al. proposed a new spatial mixing block in the ConvMixer model is the ConvMixer block, in which the spatial mixing uses only a single depth-wise convolution layer instead of self-attention or MLP token mixing.

\begin{figure}[ht!]
    \centering
    \includegraphics[width=0.8\textwidth]{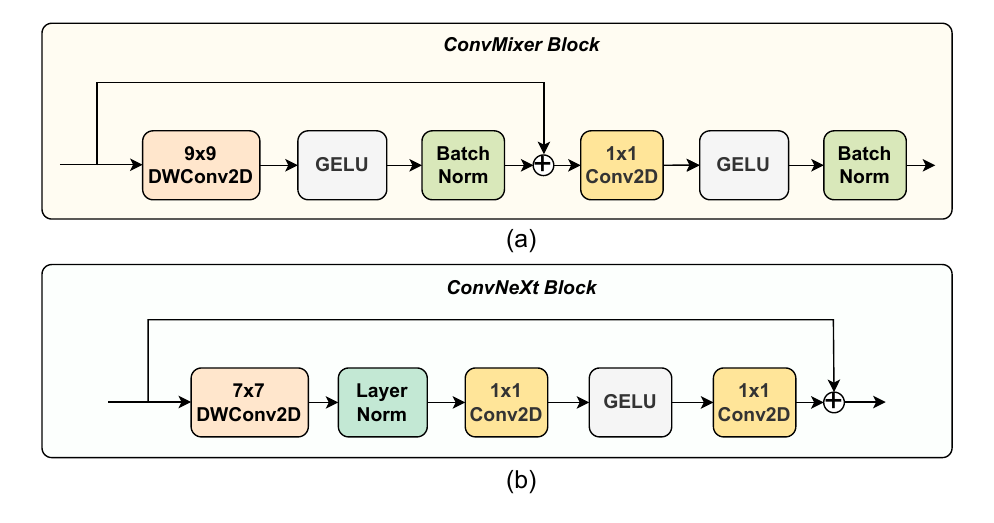}
    \caption{{General design of (a) ConvMixer block with 9x9 kernel, and (b) ConvNeXt block with 7x7 kernel}
 }
    \label{block_lge_compare}
\end{figure}

Although {of} its simple design, ConvMixer, {as shown in Fig. }\ref{block_lge_compare}a,  gives competitive results in both accuracy and number of parameters compared to ViT and MLPMixer. Yu et al. \cite{yu2022metaformer} proposed the MetaFormer model in which spatial mixing is simplified by using pooling but also gives results competitive with ViT. Liu et al. \cite{liu2022convnet} proposed the ConvNeXt model in which ConvNeXt blocks, {as illustrated in Fig. }\ref{block_lge_compare}b,  have a similar design to ConvMixer blocks but the skip connections are directly connected from the input feature map to the output feature map. ConvNeXt has completely outperformed transformer-based models on many vision tasks. In this study, we propose LGEMixer block inspired by ConvMixer block and ConvNeXt block which is capable of mixing information of pixels at different distances by using kernel of various sizes.


\subsection{Generate Context decoder}
Using connections between feature maps of equal resolution in both the encoder and decoder components is employed by Unet-based models with the aim of retaining location information. However, this approach encounters two challenges: 1) The inclusion of high-resolution feature maps in the encoder phase typically includes low-level details, while high-resolution feature maps in the decoder phase contain more high-level information. This results in an imbalance in terms of semantic information when establishing connections between them. 2) In medical images, objects often exhibit varying sizes. Therefore, to accurately segment these objects, relying solely on context information from the feature map encoder on a scale is insufficient. It becomes necessary to incorporate context information from the full-scale features map. Much previous work has attempted to address these drawbacks by using complex connections such as \cite{unet++, Unetv3} or using connection attention such as \cite{wang2022uctransnet}. However, these connections have the drawback of high computational cost while not exploiting full-scale feature map information effectively.  Generate Context Decoder (GCF) module is proposed in \cite{tran2023multi} to solve these problems. The GCF  module is inspired by self-attention mechanism of transformer model \cite{vaswani2017attention}, which uses three items: query, key and value. The key and value components are generated from the encoder's feature map, and the query component is generated from the decoder's feature map. This set of 3 components query, key, value will be combined with a lightweight attention block to create a feature map context to supplement the context information of the full-scale encoder for the decoding process. Although GCF is inspired by transformer self-attention, it has a low number of parameters and computational cost. Therefore, in this study we use GCF module to make the connection between encoder and decoder block simplified to balance computational cost and accuracy.

\subsection{Weighting In Loss Function}
A frequently employed loss function for image segmentation is the Cross entropy loss \cite{yi2004automated}, which assesses each pixel separately by comparing the class predictions (a pixel vector in term of depth) with the target vector encoded in one-hot format. The purpose of this loss function is to assess the disparities in information content between the predicted and actual image masks. Because the cross entropy loss examines each pixel vector individually and calculates an average over all pixels, equal importance is assigned to every pixel in the image. Therefore, in cases where the image masks exhibit a significant class imbalance such as medical image mask, the training process can be dominated by the most prevalent class, leading to suboptimal performance for minority classes. W-BCE \cite{WBCE} solves that problem by using a $\beta$ coefficient for positive pixels to emphasize the importance of learning pixels as objects to reduce the amount of false negatives. Another variation of W-BCE is Balanced cross entropy (B-BCE) \cite{BBCE} which uses weights for both positive pixels and negative pixels. Focal loss \cite{focalloss} is also a variant of WBCE loss where in addition to the $\beta$ weight, focal loss uses an adjustment factor based on predicted logits to downweight easy pixels and heavily penalize hard-to-predict pixels. Ronneberger et al. \cite{unet} proposed a weighting strategy for ground mask pixels by calculating a weight mask based on the function of the sum of each pixel's Euclidean distance to the object's border nearest to it and second nearest. This strategy helps the network learn better at separating the boundaries between objects. However, the disadvantage of these methods is that they do not make really good of separating the boundary area between the object and the background.

\subsection{BYOL Self-supervised Learning}
In unsupervised learning problems for computer vision tasks, learning good image feature representations is always a challenge. Some methods using contrastive learning have brought about state-of-the-art performance \cite{chen2020simple} but come with many limitations such as heavy dependence on choosing positive - negative pairs, {augmented} data sets. To solve that problem, Grill et al. \cite{byol} proposed a new approach to the self-supervised learning problem called BYOL. In this approach, they first perform data augmentation of an image in two different ways to create two different perspectives for the same image. Next, they use two networks, the online network and the target network, with architecturally identical feature extractors working together to learn the image representation. In particular, the online network inputs an augmented image to predict the target network's output with the same image as input but augmented in a different way.
In the BYOL method, the parameters of the online network will be learned by minimizing the distance between the representation vector of the online network and the target network. The parameters of the target network {are} updated according to the parameters of the online network according to the exponential moving average (EMA) \cite{ema} method. In addition, the model also uses an additional layer of predictor on the output vector of the online network to create asymmetry during the training process to prevent collapse during the learning process. In this study, we propose a training strategy inspired by BYOL called Self-embedding Representation Parallel (SeRP) to increase the robustness of the encoder in the model.

\section{Methodology}
This section presents the proposed method for medical image segmentation, namely LiteNeXt. The proposed LiteNeXt model is designed with a simplified decoder and the Self-embedding Representation Parallel (SeRP) training strategy. Unlike conventional self-supervised learning approaches that involve pretraining on large unlabeled datasets, the SeRP strategy adapts self-supervised learning principles as a data augmentation technique directly integrated into the supervised training process. In the SeRP strategy, we create two branches: the main feature extractor framework acting as the network's encoder and the target feature extractor framework. These frameworks share identical architectures for feature extraction but have different heads to handle strong geometric transformations like rotation or flipping during augmentation. The target feature extractor’s weights are updated via Exponential Moving Average (EMA) during training to enhance the robustness of the main feature extractor, which learns through backpropagation. The target feature extractor and its heads are removed during inference, reducing computational cost and parameter count. 
In addition, to handle the challenge associated with imbalance between object and background regions in common medical image segmentation tasks, we proposed a new loss function, namely Marginal Weight Loss (MWL), which incorporates a region-weighting strategy to emphasize the marginal boundaries in the loss function. This helps to effectively separate objects and background areas, particularly in complex medical image regions. The details of the proposed LiteNeXt architecture, the SeRP training strategy and the MWL are presented in the following subsections.

\subsection{The proposed model}
\begin{figure}[ht!]
\centering
\includegraphics[width=\textwidth]{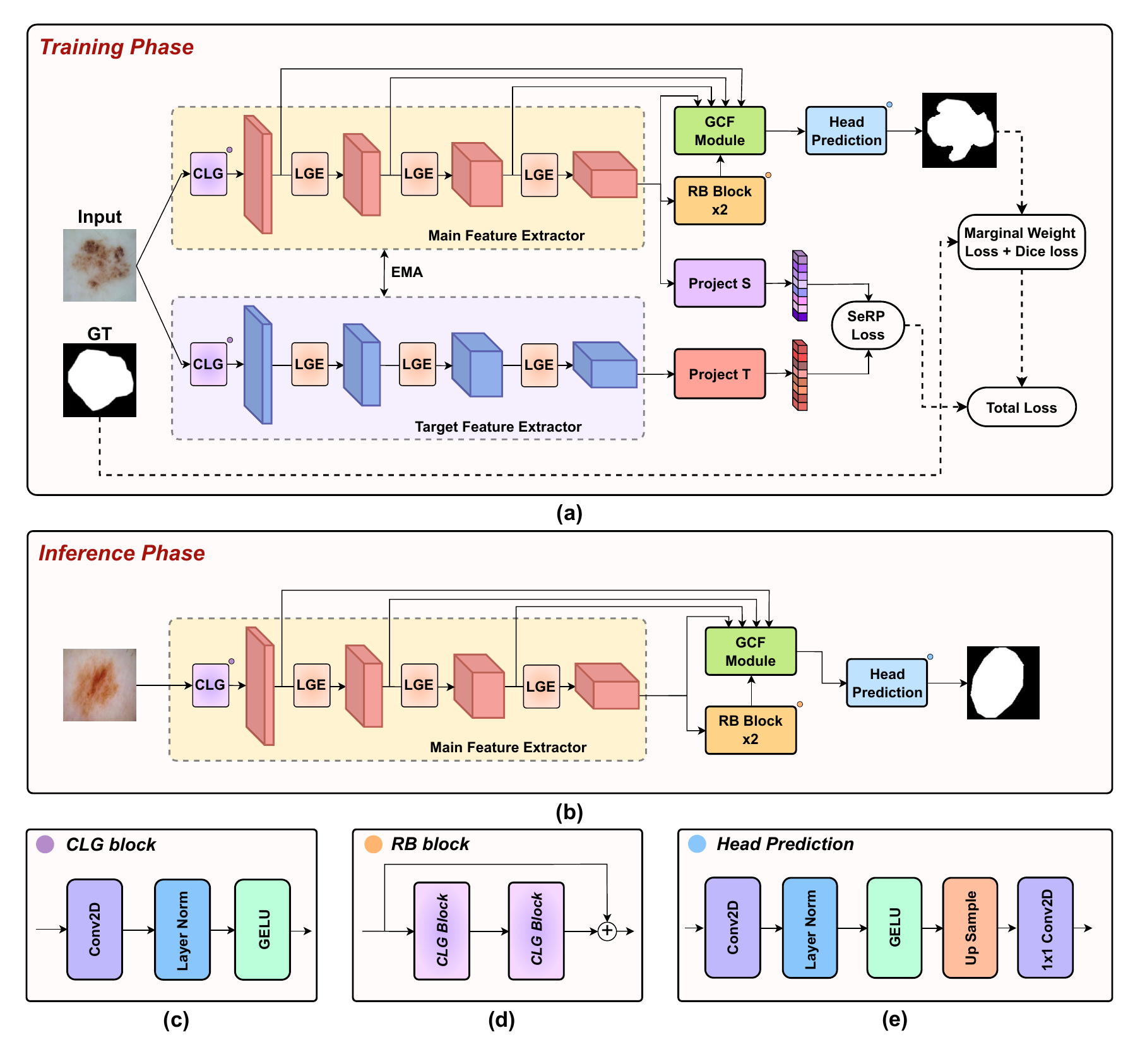}
\caption{General architecture of the proposed LiteNeXt model. (a) Overview of the training pipeline, (b) Overview of the inference phase, (c) Architecture of the CLG block, (d) Architecture of the Residual Block-abbreviated as RB block, (e) Architecture of the Head Prediction.}
\label{LiteNeXt}
\end{figure}

The overview of the proposed LiteNeXt model is illustrated in Fig. \ref{LiteNeXt}. The input image is first fed to the feature extractor and bottleneck to encode semantic information; the GCF module is also used to extract context information from the four stages of the feature extractor to add detailed information to the decoding process. {The decoder is simplified, eliminating the need for multiple layers by combining the feature map context from the GCF module with the bottleneck feature map. These combined features are then upsampled to the resolution of the mask prediction using linear interpolation.} Immediately after that, the feature map context and feature map of the bottleneck {are concatenated and further processed in a single upsampling step, reducing parameters and computational complexity.}

\subsubsection{The proposed LGEMixer Block}

Although transformer-based models {have demonstrated} superior performances over CNN-based approaches in some applications, they {are often associated with} high computational complexity. On the other hand, {recent research has shown} that some full CNN-based architectures, with low computational costs, like ConvMixer \cite{convmixer},and ConvNeXt \cite{liu2022convnet} also give outstanding performances. {Nevertheless, the ConvMixer and ConvNeXt blocks, as demonstrated in Fig.} \ref{block_lge_compare} {, only use a uniform kernel size such as 7×7 or 9×9, which may lack diversity in extracting spatial information in medical images.} In this study, we propose a lightweight feature extractor consisting of LGEMixer blocks inspired by ConvMixer and ConvNeXt blocks but with some improvements to bring higher accuracy to the segmentation task.


\begin{figure}[ht!]
    \centering
    \includegraphics[width=0.85\textwidth]{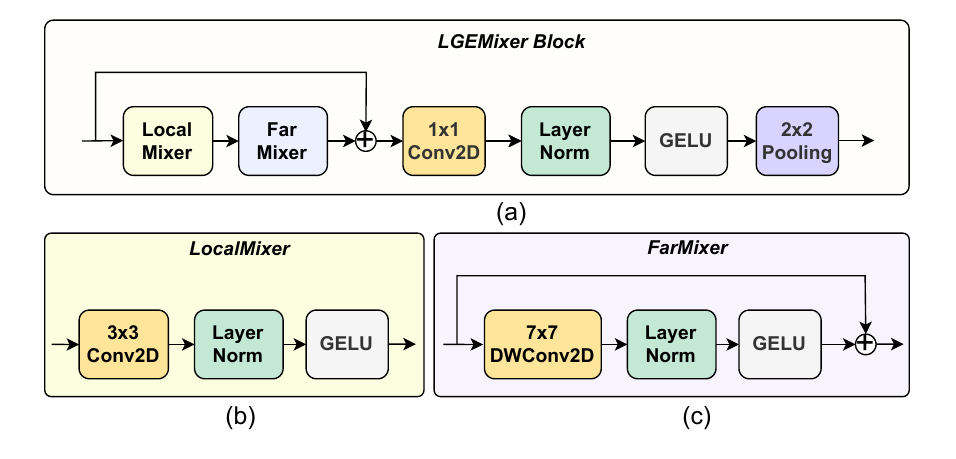}
    \caption{ Overview architectures of (a)  Our Proposed LGEMixer block, (b) LocalMixer block, (c) FarMixer block }
    \label{block_lge}
\end{figure}

The overall architecture of the LGEMixer block includes two submodules: LocalMixer block and FarMixer block. As shown in Fig.\ref{block_lge}a, the input feature map is passed through the LocalMixer block Fig.\ref{block_lge}b, which consists of a convolution layer with a kernel size of 3 followed by a Layer Normalization layer and a GELU activation function to extract detailed information about the object. Then the output feature map will be passed through the FarMixer block  Fig.\ref{block_lge}c to mix the information of pixels at a distance to expand the model's receptive field. To reduce the computational cost and number of parameters of the entire model, the FarMixer block is designed according to a residual structure, including a Depthwise Convolution layer with a large kernel size followed by a Layer normalization layer and a GELU activation function. To avoid information loss and create flexibility for the training process, a skip connection is used between the LocalMixer block and the FarMixer block. To mix the channel information, a convolution block with kernel size 1 followed by layer normalization and GELU activation is used. Finally, a max-pooling block with a pooling size of 2 is used to downsample the resolution feature map with a factor of $\frac{1}{2}$.

The detailed mathematical formula for the operation of the LGEMixer block is as follows:
\begin{equation}
    x_1 = \text{GELU}(\text{LN}(\text{Conv}_{3\times 3}(x)))
\end{equation}
\begin{equation}
    x_2 = \text{GELU}(\text{LN}(\text{DWConv}_{7\times 7}(x_1))) + x_1
\end{equation}
\begin{equation}
    x_3 = \text{GELU}(\text{LN}(\text{Conv}_{1\times 1}(x_2+x))))
\end{equation}
\begin{equation}
    y = \text{MaxPooling}_{2\times 2}(x_3)
\end{equation}
where $x\in \mathbb{R}^{C_i\times H_i\times W_i}$ is the input feature map of $ith$ layer encoder, $\text{Conv}_{3\times 3}$ is the convolution layer with kernel size of 3, $\text{DWConv}_{7\times 7}$ is the depthwise convolution with the kernel size equal to 7, $LN$ is the LayerNormalization layer, $\text{Conv}_{1\times 1}$ is the convolution layer with kernel size equal to 1.

{Compared to ConvMixer and ConvNeXt blocks that use uniform kernel sizes, the proposed LGEMixer uses two types of kernels with different sizes and reasonable design which improves the model's information extraction ability while still ensuring the requirement of a small number of parameters.} In the ablation study, the proposed LGEMixer block showed higher results than the ConvMixer block and ConvNeXt block, even though the number of parameters and GFLOPs were equivalent.

\subsubsection{Head Projector}


In the LiteNeXt model, the Head Projector is used to {facilitate} the model training process. {Specifically, Project S is applied to the main feature extractor, and Project T is applied to the target feature extractor. During inference, both Head Projectors and the target feature extractor are removed, leaving only the main feature extractor} to extract semantic information.


\begin{figure}[ht!]
    \centering
    \includegraphics[width=0.85\textwidth]{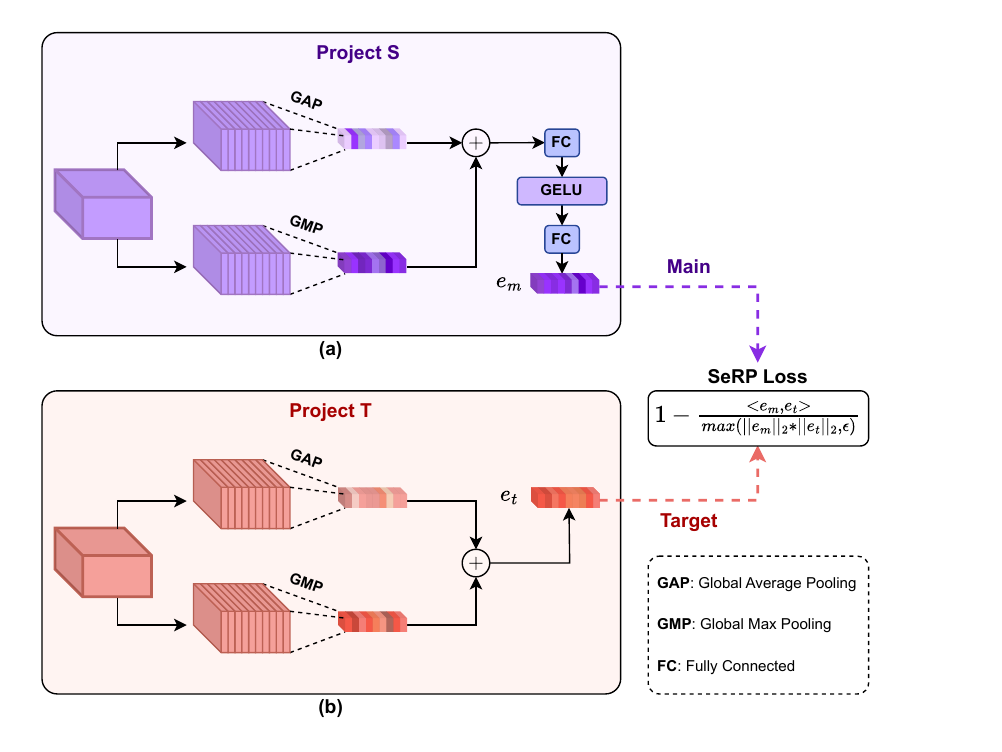}
    \caption{Overview of the Head Projector architectures, {including (a)} Project S and {(b)} Project T. {Both Project S and Project T include parallel  Global Average Pooling (GAP) and Global Max Pooling (GMP)  branches to create embedding vectors, derived from the Main Feature Extractor and Target Feature Extractor, respectively.}}
    \label{projectorh}
\end{figure}


Fig.\ref{projectorh}a illustrates Project S, {while Fig.\ref{projectorh}b illustrates Project T.} In Project T, the feature map of the target feature extractor is passed through two branches: Global Average Pooling (GAP) and Global Max Pooling (GMP), to extract both {smooth and sharp information. These two vectors are then summed to create an embedding for the input image.}

Project S, as illustrated in Fig.\ref{projectorh}a, {also includes GAP and GMP branches in parallel to create a vector embedding that synthesizes information from the main feature extractor.} {However, unlike Project T, the synthesized vector in Project S undergoes a squeeze-and-excitation layer, including MLP blocks and GELU activation, to introduce asymmetry during training.} 
The formulas for Project T and Project S are as follows:

\begin{equation}
  e_{t} = \text{GAP}(f_t) + \text{GMP}(f_t) 
\end{equation}
\begin{equation}
    v = \text{GAP}(f_m) + \text{GMP}(f_m)
\end{equation}
\begin{equation}
    e_m = \text{FC}(\text{GELU}(\text{FC}(v)))
\end{equation}
where, $f_t\in \mathbb{R}^{8C\times\frac{H}{16}\times\frac{W}{16}}$ is the feature map created from the target feature extractor, $f_m\in \mathbb{R}^{8C\times\frac{H}{16}\times\frac{W}{16}}$ is the feature map created from the main feature extractor, $e_t$, $e_m$ are the embedding vectors of projectorT and projectorS respectively, GAP and GMP respectively stand for global average pooling and global max pooling , FC represents for fully-connected layers.

\subsection{The Proposed Marginal Weight Loss (MWL) Function}
We propose a weighting strategy for three different partitions of an image based on the difficulty of segmentation. Intuitively, the background regions often occupy a larger area so segmentation of the background is relatively easier than the objects. While the background and the regions inside the desired objected can be easily segmented, the boundary between the objects and background are difficult to separate, especially in the presence of object clutter and occlusion like the cases of nuclei and histopathology images. Accordingly, we evaluate the small value of background, $w_b$ in the loss function. Next, as the partitions inside the object is commonly more difficult to segment than the background area, we weigh $w_o$ for the object to have a larger value than $ w_b$. Finally, considering the boundary region, it is most difficult for most segmentation algorithms, especially in medical images. Pixels in this area are often mistakenly recognized as background, or by contrast, the background is mistakenly recognized as an object, so the model is not able to segment the object in detail, especially in cases where there are many objects in an image and the objects are close together. Therefore in this method we will weigh the  $w_m$, with $m$ stands for margin, the largest value. In general, we have $w_m > w_o >w_b$ as the hyperparameters and to limit the search space, we set $w_m + w_o + w_b = 1$ and the search step is equal to $0.1$. \\


\begin{figure}[ht!]
    \centering
    \includegraphics[width=\textwidth]{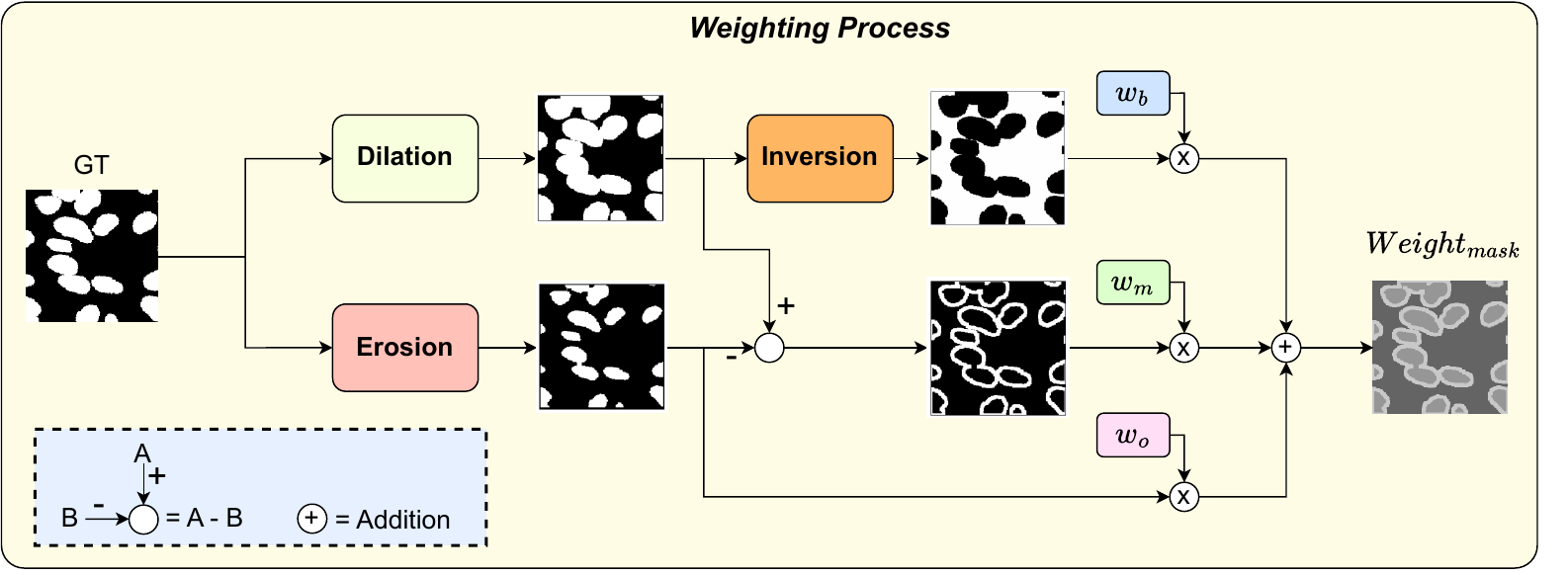}
    \caption{ Overview of the weighting process for each object partition, $w_b$ is the background weight, $w_o$ is the object weight, and $w_m$ is the marginal weight }
    \label{weightloss_margin}
\end{figure}

{The weighting process for the background, object, and margin regions is illustrated in  Fig.} \ref{weightloss_margin}. {As can be seen from this figure, two operations including erosion and dilation are directly applied on the ground truth (GT) mask by convolution with 2 kernels of the same size $k\times k$ but filled with different values following the thresholding operation. For the dilation operation, we use a kernel with values filled in by $k^2$. This ensures that if any pixel in the $k\times k$ pixel sliding window has a value of 1, the output pixel will have a value of 1. We then apply clipping between [0, 1] on the output. This guarantees that the objects in the output will have larger areas corresponding to the kernel size. Similarly, for the erosion operation, we assign a kernel value of $\frac{1}{k^2}$ to ensure that the output pixel is 1 only if all pixels in the window are 1; otherwise, the output is 0. This ensures that the objects in the output will have smaller sizes.}

{Using the outputs from the erosion and dilation operations, which produce objects with sizes different from the original objects, we then create a weighting mask, which is used throughout the training process. To determine the weights for the three regions: object, margin, and background of the weight mask, we treat them as hyperparameters and conduct experiments to find the optimal values that generalize well across common medical datasets. Due to the scalable nature of the weights, we limit the search space by choosing a step size of 0.1 within the search range [0, 1], which is sufficient.}
The detailed operation of the proposed Marginal Weight Loss (MWL) is described by pesudo-label code in Algorithm \ref{weightloss}.

\begin{algorithm}
  \caption{The algorithm to determine the weight mask for the proposed marginal weight loss (MWL)}\label{weightloss}

  \begin{algorithmic}[0]
  \State \textbf{Input:} Ground truth mask \textbf{S}, weight of each region $w_m, w_o, w_b$
  \State \textbf{Output:} Weight for ground truth mask $\textbf{Weight}_{mask}$

  \State Initialize the weights for the erosion kernel $K_e$ and the dilation kernel $K_d$ of size $k\times k$
  \State $K_e \gets \frac{1}{k^2}$\;
  \State $K_d \gets k^2$\;
  \State Calculate mask erosion $S_e$ and mask dilation $S_d$

  $S_e = K_e\odot S$

  $S_d = K_d\odot S$

  $S_e \gets \text{Floor}(S_e+ 1e-2)$

  $S_d \gets \text{Clipping}(S_d, 0, 1)$

  \State \textbf{Return:} $\textbf{Weight}_{mask} = w_m*(S_d - S_e) + w_o*S_e + w_b*(1-S_d)$
  
  \end{algorithmic}
\end{algorithm}

The loss function formula when combining weights is described as follows:
\begin{equation}
    L_{MWL}(\hat{y}_i,y_i)= -\sum_{i = 1}^{H\times W}\textbf{Weight}_{mask}(i)(y_i\log(\hat{y}_i)+(1-y_i)\log(1-\hat{y}_i))
\end{equation}
where $H, W$ are the height and width of the ground truth mask, $y_i$ is the value of the ground truth mask, $\hat{y}_i$ is the value of the predict mask.

\subsection{Proposed Self-embedding Representation Parallel (SeRP)}
It makes sense that a decent encoder would be able to derive the same representation from two somewhat different viewing angles for the same image.  In particular, in the proposed light-weight model, when the decoder branch has had most of the layers removed, the encoder branch must be able to perform well. Therefore, in this section we present a training strategy called self-embedding representation parallel (SeRP). This method is inspired by the Bootstrap Your Own Latent (BYOL) \cite{byol} model, which has achieved very good results in self-supervised learning classification tasks. To the best of our knowledge, this method is applied for the first time to train encoder robustness in a medical segmentation model end-to-end. Our method is presented as follows. \\
We use two frameworks, main framework and target framework, which have the same feature extractor architecture but have different weights and different head projectors. The target framework will provide embedding target for main framework training. The strong and weak augmentations are applied for respectively the main framework and the target framework. Specifically, for the main feature extractor branch, we apply various strong transformations to the input images to enhance data diversity during training, thereby improving the model's generalization. Meanwhile, in the target branch, we predominantly use lighter transformations to establish a reference target for the SeRP algorithm during training. For the strong augmentation in the main branch, we utilize geometric transformations such as Rotate, Flip Horizontal, and Flip Vertical, combined with pixel value transformations like ColorJitter, Gaussian Blur, and Random BrightnessContrast. The geometric transformations increase the diversity of the training data, while the pixel value transformations help the model focus on learning the intrinsic features of the objects in the images, reducing sensitivity to noise. Meanwhile, in the target branch, the objective is to create reference vectors for the main branch to facilitate learning object representations. So, in this branch, we only use geometric augmentations such as Rotate, Flip Horizontal, and Flip Vertical.

Denote the main framework and target framework as $F_m$ and $F_t$ respectively. For input image $I$, it will be augmented in two different ways to create two new images, $I_s$ and $I_w$ that stand for strong augmentation and weak augmentation images. $I_s$ and $I_w$ will be passed through the main feature extractor framework and target feature extractor framework respectively to create two embedding vectors $e_m$ and $e_t$. Finally, the loss for the SeRP will be applied to the two embeddings $e_m$ and $e_t$ as follows:

\begin{equation}
    L_{SeRP}(\theta, \tau, \phi) = 1-\frac{<e_m, e_t>}{\max{(||e_m||_2*||e_t||_2, \epsilon)}}
\end{equation}

where $\theta$ and $\tau$ are weights of the main feature extractor and projectorS in the main framework, $\phi$ is the weight of target feature extractor in the target framework and $\epsilon$ is the denominator coefficient to avoid potential divide-by-zero.
During the training process, only the weights $\theta$ and $\tau$ are learned using the back propagation algorithm, while the weights $\phi$ will be updated based on the weights $\theta$ using the Exponential Moving Average (EMA) method \cite{ema}. An overview of the full pipeline training is described in Algorithm \ref{full-pipeline}.

 \begin{algorithm}
  \caption{ Full pipeline training for LiteNeXt model}\label{full-pipeline}

  \begin{algorithmic}[0]
  \State \textbf{Input:} Training set including $N$ images and corresponding masks  $\mathbf{D} = \{x_i, y_i\}_{i=1}^{N}$
  \State \textbf{Output:} Weight $\theta$ of the main feature extractor, weight $\phi$ of the target feature extractor, weight $\tau$ of the projectorS, weight $\xi$ of the combination of modules GCF, bottleneck, head prediction.

  \State Initialize weights $\theta$, $\xi$, $\tau$, $\phi$
  \State $\theta, \xi, \tau \gets \text{He initialization}$\;
  \State $\phi \gets \theta$
  \For{$i$ = 1\textbf{ to} $N$}
       \State $I_i^s = \text{Strong Augmentation}(x_i)$; $I_i^w = \text{Weak Augmentation}(x_i)$
       \State $f_i^m= F_{\theta}(I_i^s)$;  $f_i^t = F_{\phi}(I_i^w)$
       \State $e_i^m = \textit{ProjectorS}(f_i^m)$; $e_i^t = \textit{ProjectorT}(f_i^t)$
       \State $\hat{y}_i = G_{\xi}(f_i^m)$
       \State $L_{Dice}(\hat{y}_i,y_i)=1-\frac{{2 \times \hat{y}_i\times y_i}}{{\hat{y}_i+ y_i}}$
       
       \State $L_{\text{Total}} = L_{SeRP}(e_i^m,e_i^t) + L_{MWL}(\hat{y}_i, y_i) + L_{Dice}(\hat{y}_i, y_i)$
       \State $\theta \gets \theta + lr*\nabla_{\theta}^{L_{all}}$
       \State $\tau \gets \tau + lr*\nabla_{\tau}^{L_{all}}$
       \State $\xi \gets \xi + lr*\nabla_{\xi}^{L_{all}}$
       \State $\phi \gets \alpha*\phi + (1-\alpha)*\theta$
  \EndFor
  \State \textbf{Return:} $\theta$, $\xi$
  \end{algorithmic}
\end{algorithm}

In summary, at the start of training, the main and target feature extractors in both branches are initialized with identical parameters. During training, the main feature extractor learns via backpropagation, while the target feature extractor’s weights are updated using an Exponential Moving Average of the main feature extractor’s weights. To ensure the extraction of core object features, we use a cosine loss to minimize the difference between the two feature vectors, combined with dice loss and the proposed MWL loss.
{It is worth noting that, the target feature extractor and its heads only play a role in the SeRP algorithm to enhance the robustness of the main feature extractor, acting as the model's encoder, and do not participate in generating the final prediction mask. Consequently, during inference, these components are removed from the model, reducing the parameters and computational cost.}

\section{Experiments}
\subsection{Datasets}

\subsubsection{Data Science Bowl 2018 dataset}
The 2018 Data Science Bowl dataset (Bowl2018) \cite{rashno2017fully} is a prominent dataset used in Data Science Bowl competition held in 2018. It is a large-scale dataset that attracted researchers and data scientists from around the world to tackle a specific challenge related to medical imaging. This dataset comprises a vast collection of images with segmented nuclei. These images were obtained under various conditions and demonstrate variances in cell types, magnification levels, and imaging techniques (such as brightfield or fluorescence). The main objective of the dataset is to test the capability of algorithms to generalize across these variations. To assess the efficiency of our suggested approach, we partitioned a set of 671 nuclei images with known labels into two segments: 70\% for training, 10\% for validating and 20\% for testing objective.

\subsubsection{GLAS Segmentation dataset}
The Gland Segmentation dataset (GlaS)\cite{gladseg} is specifically designed for the task of segmenting glandular structures in histopathology images. This dataset contains 165 high-resolution histopathology images, typically acquired from tissue slides stained with hematoxylin and eosin (H\&E). These images represent various tissue samples, such as breast, prostate, and colon, where glandular structures are present. Similar to the previously mentioned datasets, we divided this dataset into two subsets for the purpose of evaluating the proposed method. The training set consists of 70\% of the data, while the validating set contains 10\% of data and testing set comprises the remaining 20\%.

\subsubsection{ISIC2018 Lesion Segmentation dataset}

The ISIC 2018 dataset (ISIC2018) \cite{codella2019skin}, also known as the International Skin Imaging Collaboration 2018 dataset, is a comprehensive collection of skin images that is widely recognized and extensively utilized in the fields of dermatology and computer vision research. It was created to aid in the development and evaluate  algorithms for skin lesion analysis and classification. To evaluate our method, we utilized the Lesion Boundary Segmentation dataset from the ISIC 2018 dataset. This dataset consists of 2594 dermoscopy images of skin lesions, accompanied by expert-annotated masks from various anatomic locations and institutions. In order to assess our proposed method, we divided this dataset into three sets: 70\% for training, 10\% for validating purposes and 20\% for testing.

\subsubsection{PH2 dataset}

PH2 \cite{mendoncca2013ph} is a small dataset which was created with the intention of supporting research and providing a benchmark for evaluating segmentation and classification algorithms for dermoscopic images. It aims to enable comparative studies and facilitate advancements in these area of research. This collection comprises 200 dermoscopic images of melanocytic lesions, which are high-resolution 8-bit RGB color images with dimensions of 768x560 pixels. Among these images, there are 80 atypical nevi, 80 common nevi, and 40 melanomas. This dataset offers a diverse range of melanocytic lesions, making it a valuable resource for various research areas such as dermatology and computer vision. With no difference with all the datasets above, we also divided PH2 dataset into 3 parts with 70\% for training, 10 \% for validating and 20\% for testing purpose.

\subsubsection{Sunnybrook dataset}

 The Sunnybrook dataset \cite{radau2009evaluation} is a comprehensive compilation of cine MRI data collected from 45 patients, providing a wide array of cardiac conditions such as normal hearts, hypertrophy, heart failure with and without infarction. Originally introduced for the MICCAI 2009 challenge focused on automated left ventricle (LV) segmentation using short-axis cardiac MRI, this dataset offers meticulous manual segmentation contours by experts for the endocardium, epicardium, and papillary muscles across basal to apical slices during both end-diastole (ED) and end-systole (ES) phases. The Sunnybrook dataset holds significant value as an invaluable resource in the realm of cardiac image analysis research and advancement. These segmentation annotations are crucial for developing and testing automated algorithms for left ventricle (LV) segmentation. For this experiment, we partitioned the dataset into three distinct subsets, each including endocardium  and epicardium parts. The division was performed using a ratio of 8:1:1, allocating 80\% of the data for training, 10\% for validation, and the remaining 10\% for testing. 

\subsubsection{Lung X-ray datasets}


The lung X-ray images used in this study come from a combination of two datasets. The first dataset is the Shenzhen CXR dataset \cite{jaeger2013automatic} provided by Shenzhen No3 Hospital, which includes 326 normal CXR images and 336 CXR images with signs of pulmonary tuberculosis. The second dataset is the Montgomery dataset \cite{candemir2013lung} provided by the United States National Library of Medicine, which includes 80 normal images and 58 abnormal images related to pulmonary tuberculosis. The dataset is preprocessed by removing images without corresponding masks and all images are resized to 256x256. The final dataset consists of 566 images and 138 images from the Shenzhen and Montgomery CXR datasets, respectively. To verify the generalizability of the model on unseen datasets, we use the Shenzhen dataset for training and validation with 453 images for training and 113 images for validation, while all images of the Montgomery set are used as the testing set.



\subsection{Evaluation Metrics}
In biomedical image segmentation tasks, both Dice Similarity Coefficient (DSC) metric and Intersection over Union (IoU) are commonly used to evaluate the performance of the models. They provide a measure of how well the predicted segmentation aligns with the ground truth, with higher values indicating better agreement. The Dice coefficient measures the overlap between the predicted and the ground truths in image segmentation tasks. It ranges from 0 to 1, with 1 indicating a perfect overlap and 0 indication no overlap. The Dice coefficient is calculated as twice the intersection of the predicted mask with the ground-truth divided by their sum. The Intersection over Union (IoU) quantifies the overlap between the predicted and ground truth by calculating the ratio of their intersection to their union. It also ranges from 0 to 1, with 1 indicating a perfect match. The formulas to calculate DCS and IoU {are} determined as follows:

\begin{equation}
   \text{DSC} =  \frac{{2 \times TP}}{{2 \times TP + FP + FN + \varepsilon}}
\end{equation}

\begin{equation}
  \text{{IoU}} = \frac{{TP}}{{TP + FP + FN + \varepsilon}}
\end{equation}

{In addition to DSC and IoU, Precision and Recall are two fundamental metrics widely used to evaluate the segmentation performance. Precision and Recall are expressed as follows:}

\begin{equation} \label{equa12}
\text{Recall} = \frac{TP}{TP + FN+ \varepsilon}
\end{equation}

\begin{equation} \label{equa13}
\text{Precision} = \frac{TP}{TP + FP+ \varepsilon}
\end{equation}


where $TP$ is the number of true positives, $FP$ is the number of false positive, $FN$ is the number of false negative and $\varepsilon$ is the denominator coefficient to avoid potential divide-by-zero.

{To evaluate the segmentation performance, the F-score (also denoted as $F_1$) is used to measure the harmonic mean of precision and recall, expressed as :}
\begin{equation} \label{eq14}
    \text{F-score} = \frac{2.\text{Precision} \times \text{Recall}}{\text{Precision} + \text{Recall}}
\end{equation}

{To assess the statistical significance of the segmentation results when comparing our model with others, we calculate the $p$-value using the DSC, IoU, Precision, and Recall metrics through paired \textit{t}-tests. A $p$-value less than or equal to the threshold for significance (commonly set at 0.05) signifies that the results are statistically significant.}

\subsection{Implementation Details}
We performed LiteNeXt model experiments on the Kaggle platform, using Nvidia Tesla P100 GPU and pytorch framework version 1.12.0. In all experiments, we trained the model for 300 epochs, using the NAdam \cite{dozat2016incorporating} optimizer with a learning rate of 1e-4. To reduce overfitting, we use the $L2$ regularization method \cite{hastie2020ridge} combined with the \textit{ReduceLROnPlateau} scheduler learning rate strategy with a step size of 30 and rate drop 0.75. Data augmentation methods are also used including Rotation, HorizontalFlip, VerticalFlip, ColorJitter, RandomBrightnessContrast, GaussianBlur, RandomResizedCrop. During both training and testing, all images are resized to 256x256. The hyperparameters are set as: $w_m= 0.6, w_o=0.3, w_b= 0.1$ according to ablation study. Most of the models  used to compare with the proposed model were retrained from the official public source code with the same training settings for the proposed model. Models followed by * in the comparison tables are models with private code so the results are quoted directly from the original paper.

{To illustrate the computational complexity of the proposed method and the compared models, we present the following information in the tables for quantitative result comparison: "Param" represents the total number of parameters, while the "FLOPS" column indicates the Floating-Point Operations per Second, providing insights into the efficiency and accuracy trade-offs during both the training and inference stages. Additionally, "gpuMem" denotes the GPU memory consumption} \cite{gaogpu2000} , { and "Infer" refers to the inference latency} \cite{zhangnn2021} {of the corresponding method. The DSC, IoU, Precision, and Recall metrics are presented as the Mean score of test set and the \textit{p}-value computed on the distribution of test samples when comparing the scores by the proposed approach with those by other models. It is important to note that the F-score is computed based on the mean harmonic of Precision and Recall as Eq. \ref{eq14}, so the \textit{p}-values are not given. For better observation, the best-performing values are marked in \textbf{bold} within each metric column. Furthermore, metrics annotated with $\uparrow$ indicate that higher values represent superior segmentation performance, whereas those marked with $\downarrow$ imply that lower values are more desirable.}


\subsection{Experimental Results}

\subsubsection{Evaluation on the 2018 Data Science Bowl dataset}
\begin{figure}[ht!]
    \centering
    \includegraphics[width=\textwidth]{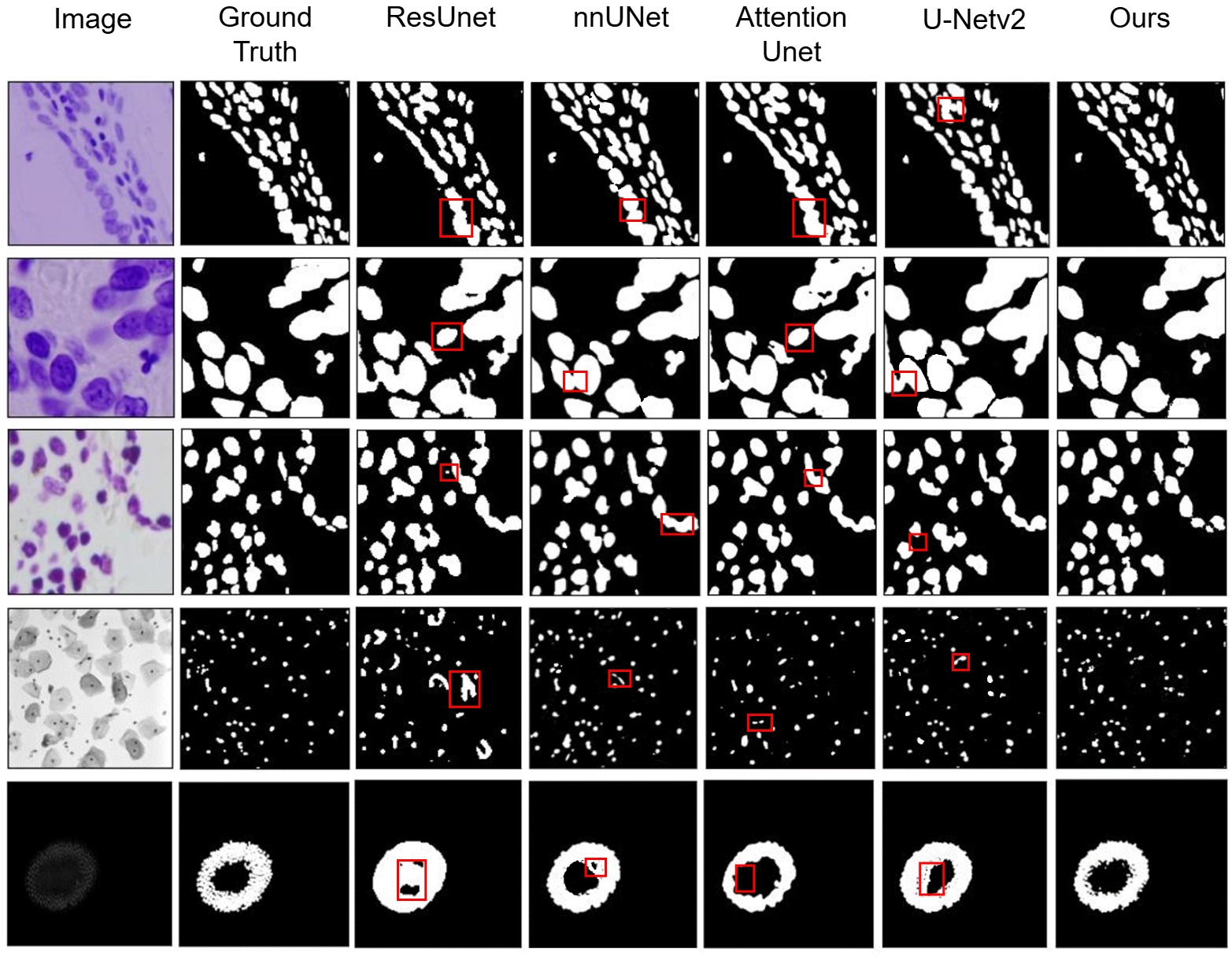}
    \caption{{Representative} visualization results of predictions on the Bowl2018 dataset. The important regions marked in red show the most difference between prediction and ground truth.}
    \label{fig:bowl_vis}
\end{figure}

In the first experiment, we verify the performance of our LiteNeXt model against different models on the Data Science Bowl 2018 dataset. To demonstrate the effectiveness of the marginal weight loss strategy in accurately separating objects, we compare the visualized images on the test set of the proposed method with different models. {Additionally, the visualization results for the top 5 in Fig. \ref{fig:bowl_vis} clearly demonstrate that our proposed model excels at distinctly segmenting cell nuclei in images, particularly in regions with overlapping or closely adjacent areas—an issue that models like nnU-Net, U-Net-V2, ResNet, and Attention U-Net struggle with.} 


Regarding quantitative assessment, the results from Table \ref{tab:bowl2018} show that our proposed model has promising performance. LiteNeXt stands out in terms of parameter count and computational cost, with only 0.71M parameters—far fewer than models such as U-Net and its variants. Even compared to lightweight models, LiteNeXt maintains an edge, using only half the number of parameters in many cases. To provide transparency in parameter and FLOP metrics, the configurations for U-Net and U-Net++ are sourced directly from their official implementations. Specifically, the reported U-Net metrics (31M parameters, 55.84G FLOPs) reflect the original paper \cite{unet}, which uses double convolutional layers at each stage. Conversely, U-Net++ metrics (9.1M parameters, 34.6G FLOPs) are based on its nested structure from \cite{unet++}, where the design reuses feature maps across layers, reducing parameter count and computational cost. However, this reuse results in higher memory consumption during inference, which LiteNeXt mitigates with its simplified architecture. Furthermore, LiteNeXt demonstrates exceptional computational efficiency with very low computational costs (0.42 GFLOPs), minimal GPU memory consumption (276 MiB), and the shortest inference latency (1.892 ms). These characteristics highlight LiteNeXt as a highly compact and computationally efficient model optimized for hardware resources, delivering extremely low latency during inference.
In terms of segmentation metrics showing the segmentation accuracy, LiteNeXt achieves outstanding performance in segmentation tasks. Two key metrics, Dice Similarity Coefficient (DSC) and Intersection over Union (IoU), demonstrate the superiority of the proposed model. The detailed metrics in Table \ref{tab:bowl2018} show that LiteNeXt achieves a DSC of 92.50\%, significantly outperforming U-Net++ (89.48\%, $p$-value = 0.0152) and lightweight models like MedT (88.47\%, $p$-value < $4.192\times10^{-5}$). Statistical significance for all metrics, with $p$-values < 0.05, confirms the reliability of LiteNeXt’s performance gains.
Similarly, LiteNeXt’s IoU metric surpasses that of complex models such as AttU-Net (83.38\%, $p$-value = 0.0324), FCN (80.88\%, $p$-value = 0.0002), nnUNet (83.00\%, $p$-value = 0.0386), and UNet (80.36\%, $p$-value = $7.624\times10^{-5}$), as well as advanced lightweight models like UNeXt (82.59\%, $p$-value = 0.0073), MedT (80.25\%, $p$-value = $1.248\times10^{-7}$), and DCSAU-Net (82.79\%, $p$-value = 0.0081). When compared to pretrained models like VMUNetV2 (80.35\%, $p$-value = $4.912\times10^{-6}$) and UNetV2 (84.21\%, $p$-value = 0.0472), LiteNeXt’s mean IoU of 86.39\% demonstrates its ability to consistently outperform state-of-the-art models across all benchmarks.
Although LiteNeXt does not consistently achieve the highest mean values for auxiliary metrics like Precision and Recall, it remains highly competitive. Its precision and recall values exhibit no significant discrepancies compared to the best-performing models, further demonstrating LiteNeXt’s ability to accurately segment cell nuclei regions while maintaining a low misclassification rate for background regions.\\
\begin{table}[ht!]
\centering
\caption{The experiment comparison between different models on the Bowl2018 dataset.}
\label{tab:bowl2018}
\resizebox{\textwidth}{!}{%
\begin{tabular}{@{}lcccccccccccccc@{}}
\toprule
\midrule
\multirow{2}{*}{\textbf{Methods}} & \multirow{2}{*}{\textbf{Params}$\downarrow$} & \multirow{2}{*}{\textbf{FLOPS}$\downarrow$} & \multirow{2}{*}{\textbf{{gpuMem}}$\downarrow$} & \multirow{2}{*}{\textbf{{Infer}}$\downarrow$} & \multicolumn{2}{c}{\textbf{DSC (\%)}$\uparrow$} & \multicolumn{2}{c}{\textbf{IoU (\%)}$\uparrow$} & \multicolumn{2}{c}{\textbf{Precision (\%)}$\uparrow$} & \multicolumn{2}{c}{\textbf{Recall (\%)}$\uparrow$} & \textbf{F-score (\%)}$\uparrow$ \\ 
\cmidrule(lr){6-7} \cmidrule(lr){8-9} \cmidrule(lr){10-11} \cmidrule(lr){12-13}
& & & & & Mean & {\textit{p}-value} & Mean & {\textit{p}-value} & Mean & {\textit{p}-value} & Mean & {\textit{p}-value}  \\ 
\midrule
AttU-Net \cite{attentionunet} & 34.87M & 66.61G & {667MiB} & {38.239ms} & {90.63} & {0.0012} & {83.38} & {0.0324} & {91.52} & {0.0485} & {90.52} & {0.0423} & {91.02} \\
FCN (VGG-based) \cite{fcn} & 134.27M & 73.45G & {915MiB} & {62.629ms} & {89.03} & {0.0024} & {80.88} & {0.0002} & {89.33} & {0.0003} & {89.81} & {0.0008} & {89.57} \\
UNeXt \cite{valanarasu2022unext} & 1.47M & 0.52G & {331MiB} & {3.928ms} & {90.17} & {0.0031} & {82.59} & {0.0073} & {90.70} & {0.0048} & {90.20} & {0.0221} & {90.45} \\
MedT \cite{valanarasu2021medical} & 1.6M & 21.24G & {423MiB} & {66.477ms} & {88.47} & {$4.192\times10^{-5}$} & {80.25} & {$1.248\times10^{-7}$} & {87.79} & {$1.293\times10^{-5}$} & {89.94} & {0.0205} & {88.86} \\
DCSAU-Net \cite{xu2023dcsau} & 2.60M & 6.91G & {491MiB} & {24.099ms} & {90.14} & {0.0076} & {82.79} & {0.0081} & \textbf{{91.93}}& {0.2085} & {89.29} & {0.0391} & {90.60} \\
UNet \cite{unet} & 31.13M & 55.84G & {655MiB} & {32.159ms} & {88.71} & {$7.624\times10^{-5}$} & {80.36} & {$2.947\times10^{-5}$} & {90.02} & {0.0212} & {88.53} & {0.0012} & {89.26} \\
UNet++ \cite{unet++} & 9.16M & 34.60G & {521MiB} & {27.167ms} & {89.48} & {0.0152} & {82.53} & {0.0054} & {90.98} & {0.0824} & {89.67} & {0.0458} & {90.32} \\
ResUnet \cite{resunet} & 13.04M & 68.07G & {675MiB} & {49.038ms} & {90.20} & {0.0305} & {82.50} & {0.0101} & {89.08} & {0.0981} & {92.08} & {0.0721} & {90.55} \\

{nnUNet} \cite{nnUnet} & {37.59M} & {0.44T} & {1015MiB} & {23768ms} & {90.36} & {0.0492} & {83.00} & {0.0386} & {88.15} & {0.0091} & \textbf{{93.64}} & {0.3921} & {90.83} \\
{U-Netv2} \cite{unetv2} & {25.15M} & {5.58G} & {417MiB} & {57.425ms} & {91.22} & {0.0758} & {84.21} & {0.0472} & {91.40} & {0.1948} & {91.60} & {0.0648} & {91.50} \\
{VM-UNetV2} \cite{vmUnetv2} & {22.77M} & {5.31G} & {451MiB} & {80.294ms} & {88.51} & {0.0008} & {80.35} & {$4.912\times10^{-6}$} & {91.67} & {0.4019} & {86.87} & {$9.543\times10^{-4}$} & {89.20} \\
\textbf{LiteNeXt} & \textbf{0.71M} & \textbf{0.42G} & \textbf{{276MiB}} & \textbf{{1.892ms}} & \textbf{92.50} & & \textbf{86.39} & & 91.85& & 93.57 & & \textbf{92.70} \\ 
\midrule
\bottomrule
\end{tabular}%
}
\end{table}
\subsubsection{Evaluation on the GlaS dataset}
\begin{figure}[ht!]
    \centering
    \includegraphics[width=\textwidth]{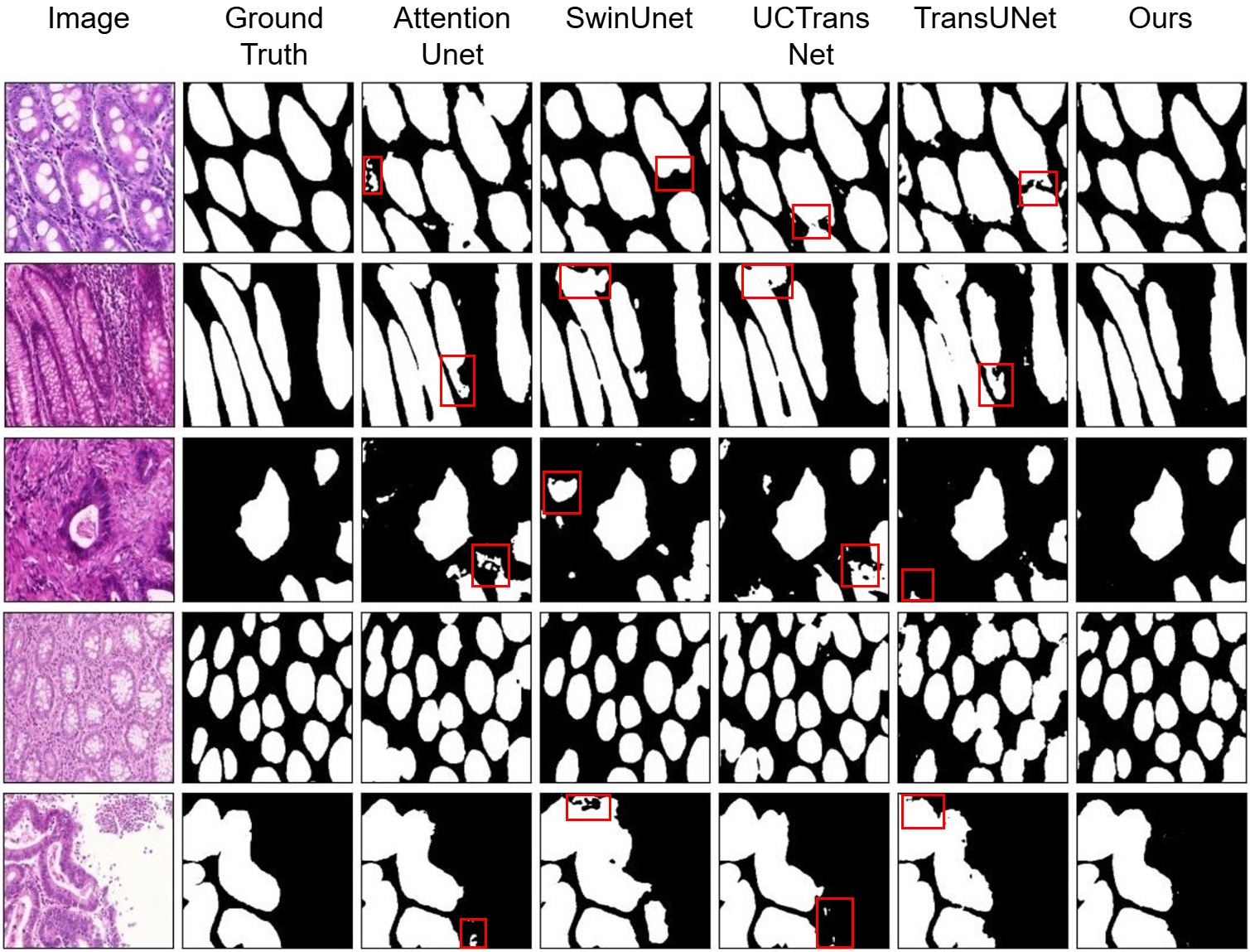}
    \caption{{Representative} visualization results of predictions on the GlaS dataset. The important regions marked in red show the most difference between prediction and ground truth.}
    \label{fig:glas}
\end{figure}

To verify its ability to work on small datasets, in this experiment, we evaluate the performance of the proposed method compared to different models on the GlaS dataset. The visualization results of {representative predictions on the test set} are shown in Fig. \ref{fig:glas}, demonstrating that our method produces prediction masks of high quality, closely resembling the ground truth. Notably, LiteNeXt effectively addresses issues such as irrelevant regions, noise leading to incorrect feature extraction, and challenges in boundary segmentation observed in other models.

Table \ref{tab:glas} provides a detailed comparison of the LiteNeXt model against other models on the GlaS dataset. {Unlike the results table for the DSB2018 dataset, this table includes more Transformer-based models. While these models often achieve high performance, they are accompanied by significant complexity, a large number of parameters, high computational costs, and longer inference times. This reduces their competitiveness compared to LiteNeXt, which excels in compactness and resource efficiency.}

In terms of performance, LiteNeXt continues to demonstrate its superiority in the two primary overlap metrics, Dice Similarity Coefficient (DSC) and Intersection over Union (IoU). {While comparative models achieve DSC values ranging from 87\% to 90\%, LiteNeXt achieves the highest value at 90.91\%. Traditional CNN-based models with a large number of parameters, such as UNet (DSC = 86.76\%, $p$-value = 0.0013), UNet++ (DSC = 88.79\%, $p$-value = 0.0102), and lightweight models like U-Lite (DSC = 86.93\%, $p$-value = 0.0027), fail to match LiteNeXt's performance. Similarly, MLP-based models such as UNeXt (DSC = 88.16\%, $p$-value = 0.0123) and Transformer-based models such as UCTransNet (DSC = 90.02\%, $p$-value = 0.0324), TransUNet (DSC = 88.93\%, $p$-value = 0.0041), and Swin-Unet (DSC = 89.67\%, $p$-value = 0.0098) as well as pretrained models like U-NetV2 (DSC = 88.60\%, $p$-value = 0.0276) and VM-UNetV2 (DSC = 88.34\%, $p$-value = 0.0121) also fail to compete. All $p$-values < 0.05 indicate a statistically significant difference in performance between LiteNeXt and the comparative models. Similarly, LiteNeXt achieves the highest average IoU compared to all evaluated models.}

Although LiteNeXt does not achieve the highest values for the auxiliary metrics Precision (91.73\%) and Recall (90.89\%), {it maintains relatively high and balanced values. Some models, such as UNet++ (Precision = 91.96\%, $p$-value = 0.2754 > 0.05), nnUNet (Recall = 90.43\%, $p$-value = 0.0891 > 0.05), VM-UNetV2 (Recall = 87.09\%, $p$-value = 0.0943 > 0.05), ResUNet (Recall = 89.77\%, $p$-value = 0.0691 > 0.05), and UNeXt (Recall = 91.13\%, $p$-value = 0.1849 > 0.05), exhibit similar values for Precision or Recall but do not achieve statistically significant differences compared to LiteNeXt.}

{In summary, when evaluated on a smaller dataset like GlaS, LiteNeXt continues to assert its superiority, leading in overlap metrics (DSC and IoU). Although Precision and Recall are not the highest, they demonstrate stability and balance, affirming LiteNeXt's strong predictive ability in identifying objects in images. The model not only maintains high performance but also ensures compactness, computational efficiency, and optimal hardware resource utilization.}



\begin{table}[ht!]
\centering
\caption{The experiment comparison between different models on the GlaS dataset.}
\label{tab:glas}
\resizebox{\textwidth}{!}{%
\begin{tabular}{@{}lcccccccccccccccc@{}}
\toprule
\multirow{2}{*}{\textbf{Methods}} & \multirow{2}{*}{\textbf{Params}$\downarrow$} & \multirow{2}{*}{\textbf{FLOPS}$\downarrow$} & \multirow{2}{*}{\textbf{{gpuMem}}$\downarrow$} & \multirow{2}{*}{\textbf{{Infer}}$\downarrow$} & \multicolumn{2}{c}{\textbf{DSC (\%)$\uparrow$}} & \multicolumn{2}{c}{\textbf{IoU (\%)$\uparrow$}} & \multicolumn{2}{c}{\textbf{Precision (\%)$\uparrow$}} & \multicolumn{2}{c}{\textbf{Recall (\%)$\uparrow$}} & \textbf{F-score (\%)$\uparrow$} \\ 
\cmidrule(lr){6-7} \cmidrule(lr){8-9} \cmidrule(lr){10-11} \cmidrule(lr){12-13}
& & & & & Mean & {\textit{p}-value} & Mean & {\textit{p}-value} & Mean & {\textit{p}-value} & Mean & {\textit{p}-value}  \\ 
\midrule
UCTransNet \cite{wang2022uctransnet} & 65.6M & 63.20G & {741MiB} & {84.823ms} & 90.02 & {0.0324} & 82.19 & {0.0237} & 90.11 & {0.0132} & \textbf{91.67} & {0.2176} & 90.88 \\
TransUNet \cite{transunet} & 105.32M & 38.52G & {825MiB} & {44.746ms} & 88.93 & {0.0041} & 80.88 & {0.0192} & 88.99 & {0.0248} & 91.38 & {0.3283} & 90.17 \\
Swin-Unet \cite{swinunet} & 41.38M & 150.72G & {529MiB} & {20.930ms} & 89.67 & {0.0098} & 81.98 & {0.0411} & 89.15 & {0.0014} & 89.96 & {0.0312} & 89.55 \\
AttU-Net \cite{attentionunet} & 34.87M & 66.61G & {667MiB} & {38.239ms} & 89.37 & {0.0075} & 81.63 & {0.0186} & 89.35 & {0.0078} & 90.73 & {0.0491} & 90.03 \\
UNet \cite{unet} & 31.13M & 55.84G & {655MiB} & {32.159ms} & 86.76 & {0.0013} & 77.61 & {$4.372\times10^{-5}$} & 88.51 & {0.0011} & 86.89 & {$3.143\times10^{-5}$} & 87.69 \\
UNet++ \cite{unet++} & 9.16M & 34.60G & {521MiB} & {27.167ms} & 88.79 & {0.0102} & 80.73 & {0.0057} & \textbf{91.96} & {0.2754} & 86.98 & {0.0002} & 89.40 \\
ResUnet \cite{resunet} & 13.04M & 68.07G & {675MiB} & {49.038ms} & 88.33 & {0.0203} & 79.97 & {0.0023} & 88.56 & {0.0009} & 89.77 & {0.0691} & 89.16 \\
U-Lite \cite{dinh20231m} & 0.88M & 1.42G & {343MiB} & {4.708ms} & 86.93 & {0.0027} & 77.98 & {$3.324\times10^{-4}$} & 87.21 & {$8.129\times10^{-5}$} & 87.74 & {0.0217} & 87.47 \\
UNeXt \cite{valanarasu2022unext} & 1.47M & 0.52G & {331MiB} & {3.928ms} & 88.16 & {0.0123} & 79.70 & {0.0005} & 86.63 & {$3.374\times10^{-6}$} & 91.13 & {0.1849} & 88.82 \\
MedT \cite{valanarasu2021medical} & 1.60M & 21.24G & {423MiB} & {66.477ms} & 87.61 & {0.0098} & 78.77 & {0.0235} & 87.89 & {0.0012} & 88.82 & {0.0084} & 88.35 \\

{nnUNet} \cite{nnUnet} & {37.59M} & {0.44T} & {1015MiB} & {23768ms} & {87.96} & {0.0078} & {79.45} & {0.0297} & {87.37} & {0.0023} & {90.43} & {0.0891} & {88.94} \\
{U-Netv2} \cite{unetv2} & {25.15M} & {5.58G} & {417MiB} & {57.425ms} & {88.60} & {0.0276} & {80.03} & {0.0354} & {88.99} & {0.0234} & {88.78} & {0.0432} & {88.88} \\
{VM-UNetV2} \cite{vmUnetv2} & {22.77M} & {5.31G} & {451MiB} & {80.294ms} & {88.34} & {0.0121} & {79.48} & {0.0293} & {90.20} & {0.0349} & {87.09} & {0.0943} & {88.63} \\
\textbf{LiteNeXt} & \textbf{0.71M} & \textbf{0.42G} & \textbf{{276MiB}} & \textbf{{1.892ms}} & \textbf{90.91} & & \textbf{84.15} & & 91.73 & & 90.89 & & \textbf{91.31} \\ 
\bottomrule
\end{tabular}%
}
\end{table}

\subsubsection{Evaluation on the ISIC 2018 dataset}

\begin{figure}[ht!]
    \centering
    \includegraphics[width=\textwidth]{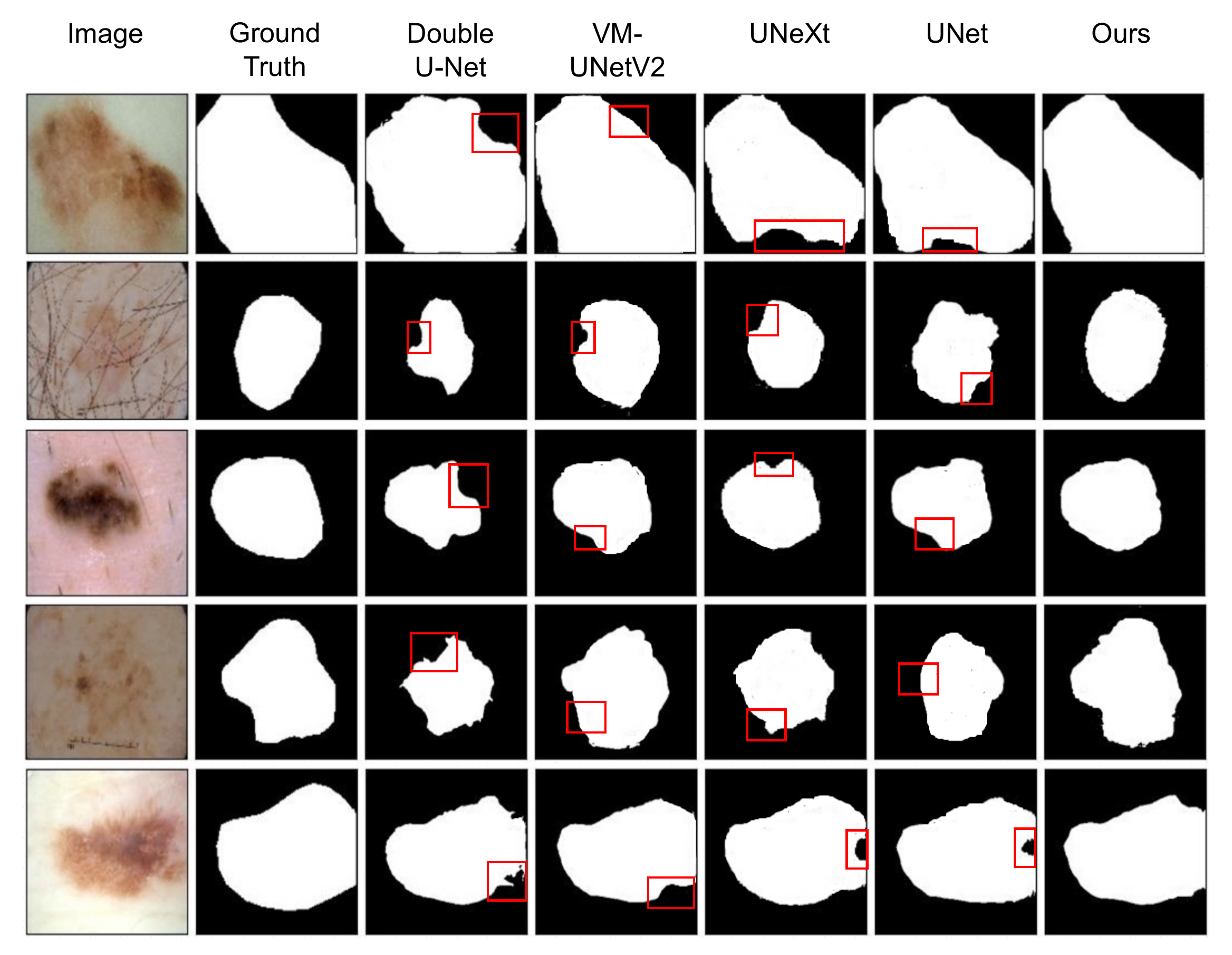}
    \caption{{Representative} visualization results of the predictions on the ISIC2018 dataset. The
important regions marked in red show the most difference between prediction and ground
truth.}
    \label{fig:isic2018}
\end{figure}

Next, to further verify the effectiveness of the proposed method in medical image segmentation, we evaluate its performance on the ISIC2018 dataset. Visualization results from Fig.\ref{fig:isic2018} demonstrate that our proposed method produces predictions that closely match the ground truth better than other methods. Notably, in cases where skin damage boundaries are unclear, LiteNeXt exhibits accurate segmentation, whereas models like VM-UNetV2 and UNet often suffer from over-segmentation or under-segmentation.

The comparison results between the LiteNeXt model combined with Marginal Weight loss and self-embedding are presented in Table \ref{tab:isic2018}. Several other models, including MSRF-Net, DeepLabv3, and Double UNet, are also included in this experiment. However, all these models lag behind LiteNeXt in terms of compactness and memory efficiency.

In terms of performance, LiteNeXt continues to excel in the primary metrics, DSC and IoU. LiteNeXt achieves the highest mean DSC value (90.52\%), outperforming all evaluated models. The \textit{t}-test results indicate a statistically significant difference ($p$-value < 0.05) between LiteNeXt and other models, except for VM-UNetV2 (DSC = 89.73\%, $p$-value = 0.0821 > 0.05), where no significant difference is observed. Similarly, LiteNeXt achieves the highest mean IoU, with statistically significant differences compared to most models except VM-UNetV2 (IoU = 83.08\%, $p$-value = 0.0824).

LiteNeXt does not achieve the highest values for the auxiliary metrics Precision and Recall but maintains good and balanced levels for both. For example, the model with the highest precision is Double UNet (Precision = 94.59\%, $p$-value = 0.4532 > 0.05), but its recall value is relatively low (recall = 87.80\%, $p$-value = $2.232\times10^{-5}$ < 0.05), indicating that Double UNet misses many pixels belonging to the region of the injury. In contrast, the model with the highest recall is nnUNet (Recall = 92.68\%, $p$-value = 0.1126 > 0.05), but its precision is unbalanced, at only 86.01\% ($p$-value = $2.977\times10^{-12}$ < 0.05). Furthermore, nnUNet exhibits high parameter counts, computational complexity, high GPU consumption, and extremely long inference latency (23768ms).

From these results, LiteNeXt, while not leading in Precision and Recall, achieves good and balanced values in both metrics. This affirms LiteNeXt as an effective segmentation model, combining superior compactness, segmentation performance, low computational cost, and optimal memory utilization.

\begin{table}[ht!]
\centering
\caption{The experiment comparison between different models on the ISIC2018 dataset.}
\label{tab:isic2018}
\resizebox{\textwidth}{!}{%
\begin{tabular}{@{}lcccccccccccccccc@{}}
\toprule
\multirow{2}{*}{\textbf{Methods}} & \multirow{2}{*}{\textbf{Params}$\downarrow$} & \multirow{2}{*}{\textbf{FLOPS}$\downarrow$} & \multirow{2}{*}{\textbf{{gpuMem}}$\downarrow$} & \multirow{2}{*}{\textbf{{Infer}}$\downarrow$} & \multicolumn{2}{c}{\textbf{DSC (\%)$\uparrow$}} & \multicolumn{2}{c}{\textbf{IoU (\%)$\uparrow$}} & \multicolumn{2}{c}{\textbf{Precision (\%)$\uparrow$}} & \multicolumn{2}{c}{\textbf{Recall (\%)$\uparrow$}} & \textbf{F-score (\%)$\uparrow$} \\ 
\cmidrule(lr){6-7} \cmidrule(lr){8-9} \cmidrule(lr){10-11} \cmidrule(lr){12-13}
& & & & & Mean & {\textit{p}-value} & Mean & {\textit{p}-value} & Mean & {\textit{p}-value} & Mean & {\textit{p}-value}  \\ 
\midrule
AttU-Net \cite{attentionunet} & \textcolor{black}{34.87M} & \textcolor{black}{66.61G} & {667MiB} & {38.239ms} & \textcolor{black}{88.13} & {0.0352} & \textcolor{black}{81.58} & {0.0294} & \textcolor{black}{89.24} & {0.0054} & \textcolor{black}{88.76} & {0.0021} & {88.99} \\
UNet 3+ \cite{Unetv3} & {26.98M} & {56.87G} & {905MiB} & {110.539ms} & {88.71} & {0.0007} & {82.56} & {0.0097} & {88.19} & {0.0018} & {90.34} & {0.0105} & {89.27} \\
MSRF-Net \cite{msrfnet} & \textcolor{black}{22.50M} & \textcolor{black}{83.77G} & {651MiB} & {121.714ms} & \textcolor{black}{88.14} & {0.0319} & \textcolor{black}{82.12} & {0.0420} & \textcolor{black}{93.48} & {0.3381} & \textcolor{black}{88.93} & {0.0064} & {90.68} \\
Double U-Net \cite{doubleunet} & \textcolor{black}{29.28M} & \textcolor{black}{53.81G} & {585MiB} & {49.247ms} & \textcolor{black}{89.01} & {0.0416} & \textcolor{black}{82.18} & {0.0389} & \textbf{\textcolor{black}{94.59}}& {0.4532} & \textcolor{black}{87.80} & {$2.232\times10^{-5}$} & {91.07} \\
DeepLabv3 \cite{deeplabv3} & \textcolor{black}{11.03M} & \textcolor{black}{2.48G} & {351MiB} & {18.194ms} & \textcolor{black}{87.81} & {0.0011} & \textcolor{black}{81.22} & {0.0209} & \textcolor{black}{89.91} & {0.0156} & \textcolor{black}{88.76} & {0.0112} & {89.33} \\
TransUNet \cite{transunet} & \textcolor{black}{105.32M} & \textcolor{black}{38.52G} & {825MiB} & {44.746ms} & \textcolor{black}{88.99} & {0.0031} & \textcolor{black}{82.07} & {0.0276} & \textcolor{black}{93.76} & {0.0594} & \textcolor{black}{88.89} & {0.0045} & {91.26} \\
Swin-Unet \cite{swinunet} & \textcolor{black}{41.38M} & \textcolor{black}{150.72G} & {529MiB} & {20.930ms} & \textcolor{black}{88.70} & {0.0162} & \textcolor{black}{82.62} & {0.0316} & \textcolor{black}{93.48} & {0.1432} & \textcolor{black}{88.93} & {0.0038} & {91.15} \\
UNeXt \cite{valanarasu2022unext} & \textcolor{black}{1.47M} & \textcolor{black}{0.52G} & {331MiB} & {3.928ms} & {89.68} & {0.0027} & {82.79} & {0.0193} & {91.85} & {0.0385} & {90.38} & {0.0009} & {89.97} \\
MedT \cite{valanarasu2021medical} & \textcolor{black}{1.6M} & \textcolor{black}{21.24G} & {423MiB} & {66.477ms} & \textcolor{black}{87.16} & {$7.412\times10^{-5}$} & \textcolor{black}{81.12} & {0.0017} & \textcolor{black}{88.46} & {$5.923\times10^{-7}$} & \textcolor{black}{87.92} & {$1.164\times10^{-5}$} & {88.15} \\
DCSAU-Net \cite{xu2023dcsau} & \textcolor{black}{2.60M} & \textcolor{black}{6.91G} & {491MiB} & {24.099ms} & \textcolor{black}{88.48} & {0.0477} & \textcolor{black}{82.32} & {0.0454} & \textcolor{black}{94.50} & {0.3215} & \textcolor{black}{87.31} & {0.0008} & {90.76} \\
UNet \cite{unet} & \textcolor{black}{31.13M} & \textcolor{black}{55.84G} & {655MiB} & {32.159ms} & {89.57} & {$8.135\times10^{-6}$} & {82.29} & {0.0048} & {91.01} & {0.0422} & {90.26} & {$4.129\times10^{-6}$} & {89.78} \\
UNet++ \cite{unet++} & \textcolor{black}{9.16M} & \textcolor{black}{34.60G} & {521MiB} & {27.167ms} & {88.02} & {0.0293} & {81.28} & {0.0353} & {91.74} & {0.0511} & {86.81} & {0.0006} & {89.21} \\
ResUnet \cite{resunet} & \textcolor{black}{13.04M} & \textcolor{black}{68.07G} & {675MiB} & {49.038ms} & {88.96} & {0.0412} & {81.36} & {0.0131} & {93.91} & {0.0984} & {86.51} & {$6.563\times10^{-6}$} & {89.86} \\
{nnUNet} \cite{nnUnet} & {37.59M} & {0.44T} & {1015MiB} & {23768ms} & {87.60} & {0.0009} & {79.46} & {0.0193} & {86.01} & {$2.977\times10^{-12}$} & \textbf{{92.68} }& {0.1126} & {88.92} \\
{U-Netv2} \cite{unetv2} & {25.15M} & {5.58G} & {417MiB} & {57.425ms} & {88.97} & {0.0446} & {81.89} & {0.0368} & {90.93} & {0.0375} & {90.05} & {0.1376} & {89.96} \\
{VM-UNetV2} \cite{vmUnetv2} & {22.77M} & {5.31G} & {451MiB} & {80.294ms} & {89.73} & {0.0821} & {83.08} & {0.0824} & {91.84} & {0.0179} & {90.43} & {0.0294} & {89.91} \\
\textbf{LiteNeXt} & \textbf{0.71M} & \textbf{0.42G} & \textbf{{276MiB}} & \textbf{{1.892ms}} & \textbf{90.52} & & \textbf{83.93} & & 92.00 & & 91.29 & & \textbf{91.64} \\ 
\bottomrule
\end{tabular}%
}
\end{table}

\subsubsection{Evaluation on the PH2 data}

\begin{figure}[ht!]
    \centering
    \includegraphics[width=\textwidth]{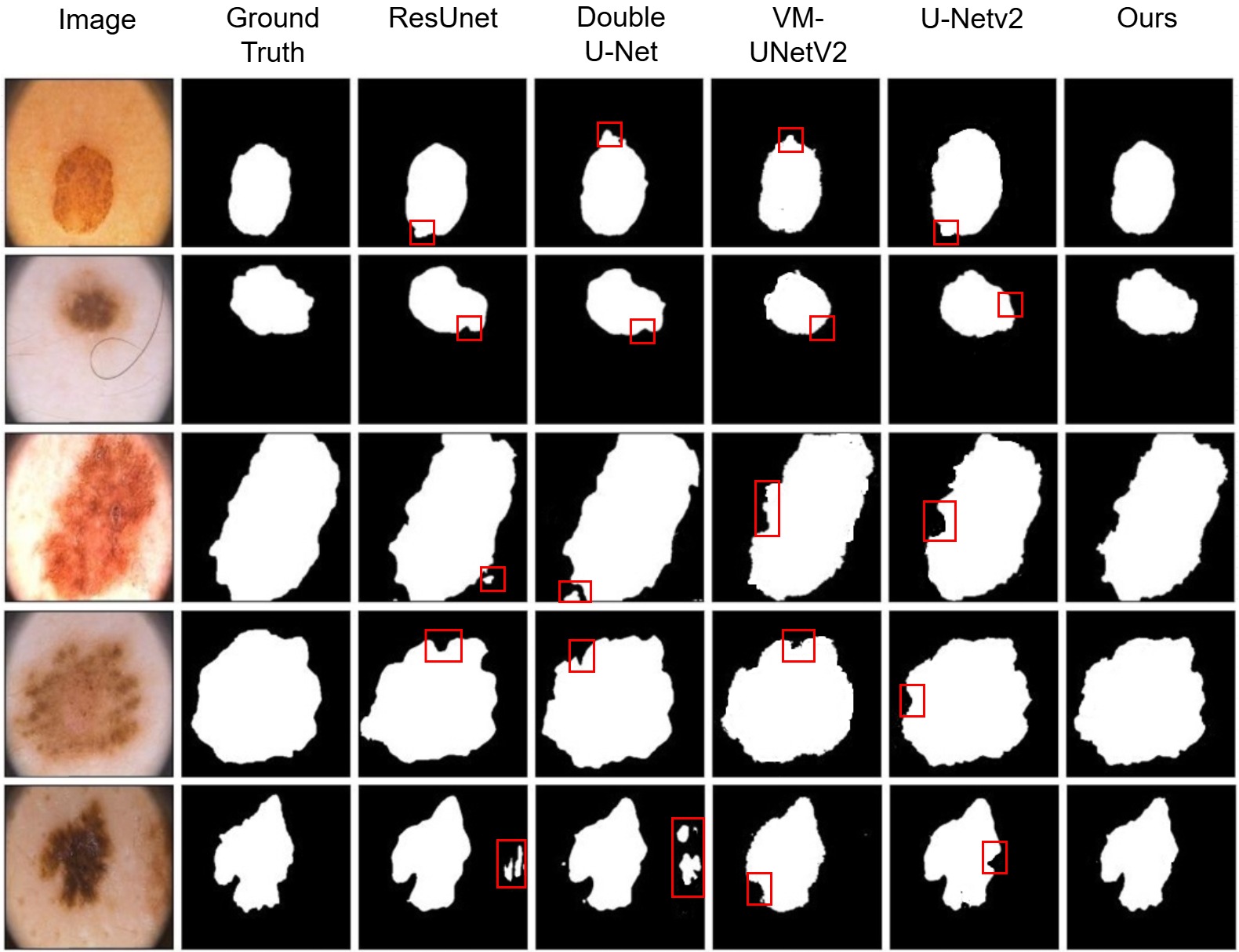}
    \caption{{Representative} visualization results of predictions on the PH2 dataset. The
important regions marked in red show the most difference between prediction and ground
truth.}
    \label{fig:ph2}
\end{figure}

To further evaluate the performance of the proposed approach for small datasets, in this experiment, we compare the proposed method with different models on the PH2 data. Fig. \ref{fig:ph2} {shows the representative visualization results of predictions on the test set. It is evident that the segmentation predictions by our method align better with ground truths compared to the others. Furthermore, similar to the ISIC18 dataset, LiteNeXt demonstrates superior segmentation performance in handling unclear boundary regions compared to previous models.}

{The detailed comparison results between LiteNeXt and other models are presented in Table \ref{tab:ph2}. Some models marked (*) are private implementations. Overall, LiteNeXt continues to assert its superiority with a compact design, simple computations, low GPU memory consumption, and minimal inference latency, while maintaining high segmentation performance.}

{The two primary metrics, DSC and IoU, achieve the highest mean values among all evaluated models (DSC = 95.18\%, IoU = 90.98\%). For DSC, two models showed no statistically significant difference compared to LiteNeXt: ResUNet (DSC = 93.98\%, $p$-value = 0.0674) and U-NetV2 (DSC = 94.83\%, $p$-value = 0.0894). However, these $p$-values are still relatively small, indicating the competitive performance of LiteNeXt. Similarly, for IoU, Double UNet, ResUNet, and U-NetV2 also showed no clear statistical differences compared to LiteNeXt. Nonetheless, LiteNeXt outperforms these models in other aspects, including parameter count, computation cost, GPU memory consumption, and inference latency, reinforcing its overall efficiency.}

{As with previously evaluated datasets, LiteNeXt’s auxiliary metrics Precision and Recall achieve good and highly balanced levels. Although LiteNeXt does not lead in individual metrics, both Precision and Recall are high, demonstrating stability and accurate segmentation, while minimizing confusion between object and background regions.}

\begin{table}[ht!]
\centering
\caption{The experiment comparison between different models on the PH2 dataset.}
\label{tab:ph2}
\resizebox{\textwidth}{!}{%
\begin{tabular}{@{}lccccccccccccccc@{}}
\toprule
\multirow{2}{*}{\textbf{Methods}} & \multirow{2}{*}{\textbf{Params}$\downarrow$} & \multirow{2}{*}{\textbf{FLOPS}$\downarrow$} & \multirow{2}{*}{\textbf{{gpuMem}}$\downarrow$} & \multirow{2}{*}{\textbf{{Infer}}$\downarrow$} & \multicolumn{2}{c}{\textbf{DSC (\%)$\uparrow$}} & \multicolumn{2}{c}{\textbf{IoU (\%)$\uparrow$}} & \multicolumn{2}{c}{\textbf{Precision (\%)$\uparrow$}} & \multicolumn{2}{c}{\textbf{Recall (\%)$\uparrow$}} & \textbf{F-score (\%)$\uparrow$} \\ 
\cmidrule(lr){6-7} \cmidrule(lr){8-9} \cmidrule(lr){10-11} \cmidrule(lr){12-13}
& & & & & Mean & {\textit{p}-value} & Mean & {\textit{p}-value} & Mean & {\textit{p}-value} & Mean & {\textit{p}-value}  \\ 
\midrule
Double U-Net \cite{doubleunet} & 29.28M & 53.81G & {585MiB} & {49.247ms} & {93.89} & {0.0494} & {89.28} & {0.0898} & {92.33} & {0.1575} & {\textbf{96.75}} & {0.1418} & {94.49} \\
SegNet \cite{badrinarayanan2017segnet} & 29.44M & 20.12G & {472MiB} & {27.631ms} & {89.88} & {0.0038} & {82.78} & {0.0015} & {90.97} & {0.0604} & {91.41} & {0.0482} & {91.19} \\
UNeXt \cite{valanarasu2022unext} & 1.47M & 0.52G & {331MiB} & {3.928ms} & {90.37} & {$1.845\times10^{-3}$} & {83.25} & {$6.102\times10^{-4}$} & {90.55} & {0.0224} & {92.35} & {0.1269} & {91.44} \\
MedT \cite{valanarasu2021medical} & 1.6M & 21.24G & {423MiB} & {66.477ms} & {93.10} & {0.0475} & {87.89} & {0.0472} & {93.71} & {0.3881} & {93.99} & {0.4166} & {93.85} \\
AttU-Net \cite{attentionunet} & 34.87M & 66.61G & {667MiB} & {38.239ms} & {91.20} & {$1.490\times10^{-3}$} & {84.36} & {$8.043\times10^{-4}$} & {90.63} & {0.0247} & {93.22} & {0.1424} & {91.91} \\
UNet \cite{unet} & 31.13M & 55.84G & {655MiB} & {32.159ms} & {92.46} & {0.0466} & {86.70} & {0.0394} & {93.81} & {0.3290} & {92.46} & {0.1371} & {93.13} \\

{UNet++} \cite{unet++} & {9.16M} & {34.60G} & {521MiB} & {27.167ms} & {92.21} & {0.0067} & {86.16} & {0.0063} & {90.54} & {0.0122} & {95.16} & {0.7896} & {92.79} \\
ResUnet \cite{resunet} & 13.04M & 68.07G & {675MiB} & {49.038ms} & {93.98} & {0.0674} & {88.97} & {0.0789} & {95.17} & {0.8826} & {93.60} & {0.1594} & {94.37} \\
DSnet \cite{hasan2020dsnet} & 10.14M & 13.44G & {417MiB} & {20.038ms} & {92.98} & {0.0425} & {87.35} & {0.0478} & {92.94} & {0.0162} & {94.15} & {0.4588} & {93.54} \\
Goyal et al. \cite{goyal2019skin} & - & - & - & - & 90.70 & - & 83.92 & - & 93.24 & - & - & - & - \\
Ozturk et al. \cite{ozturk2020skin} & - & - & - & - & 93.02 & - & 87.10 & - & \textbf{96.92} & - & - & - & - \\
Unver et al. \cite{unver2019skin} & - & - & - & - & 88.13& - & 79.54 & - & 83.63 & - & - & - & - \\
{nnUNet} \cite{nnUnet} & {37.59M} & {0.44G} & {1015MiB} & {23768ms} & {92.33} & {0.0438} & {86.50} & {0.0375} & {91.93} & {0.1341} & {94.32} & {0.4274} & {93.11} \\
{U-Netv2} \cite{unetv2} & {25.15M} & {5.58G} & {417MiB} & {57.425ms} & {94.83} & {0.0894} & {90.42} & {0.0872} & {94.84} & {0.6987} & {95.12} & {0.8854} & {94.98} \\
{VM-UNetV2} \cite{vmUnetv2} & {22.77M} & {5.31G} & {451MiB} & {80.294ms} & {93.37} & {0.0165} & {88.20} & {0.0425} & {89.64} & {0.4014} & {96.32} & {0.4148} & {92.86} \\
\textbf{LiteNeXt} & \textbf{0.71M} & \textbf{0.42G} & \textbf{{276MiB}} & \textbf{{1.892ms}} & \textbf{95.18} & & \textbf{90.98} & & {95.42} & & {95.33} & & \textbf{95.37} \\ 
\bottomrule
\end{tabular}
}
\end{table}

\subsubsection{Evaluation on the Sunnybrook (SB) data}

\begin{figure}[ht!]
    \centering
    \subfloat[Endo visualization]{\includegraphics[width=\textwidth]{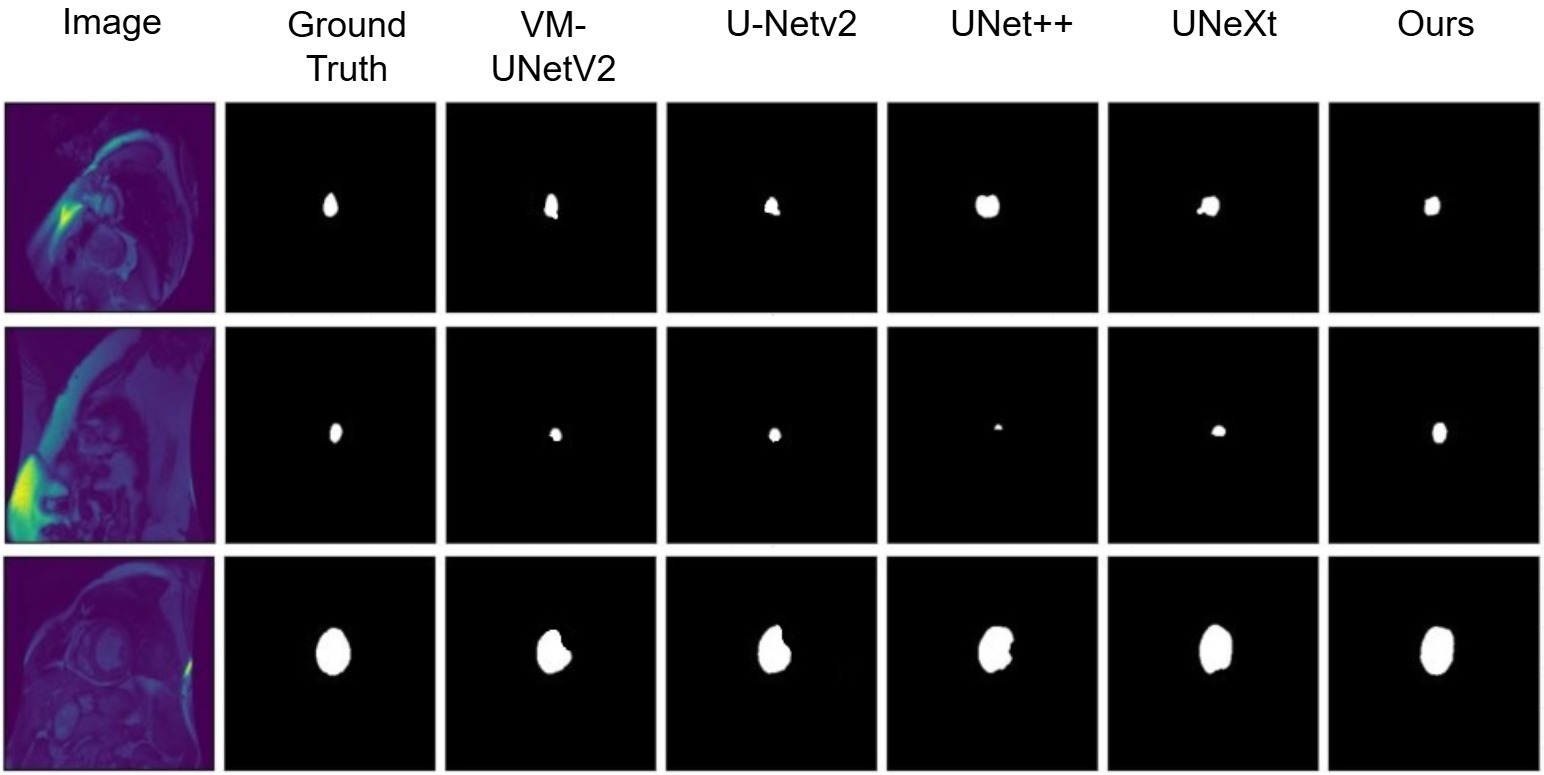}}
    \hfill
    \subfloat[Epi visualization]{\includegraphics[width=\textwidth]{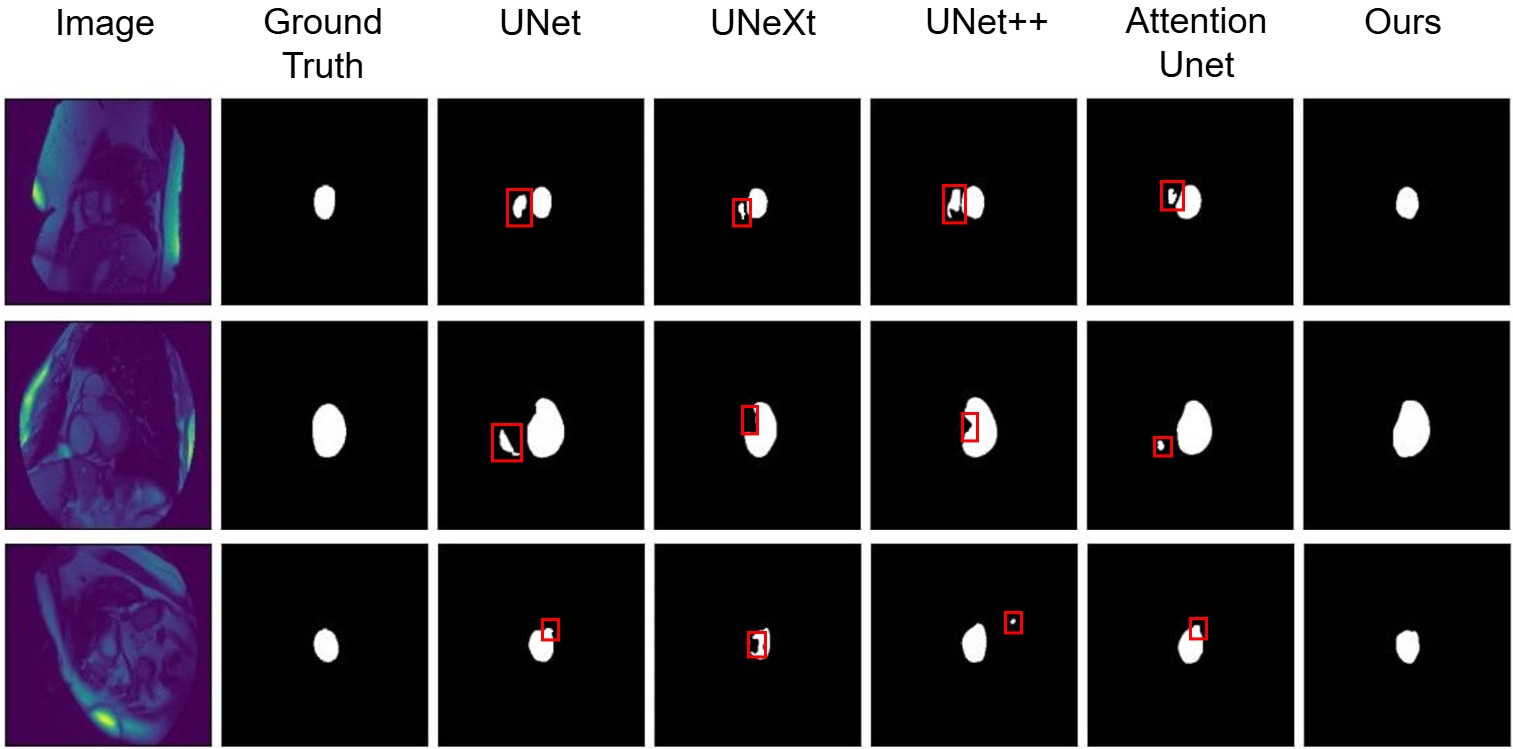}}

    \caption{{Representative} visualization results of prediction models, (a) Visualization results of Endocardium (Endo), (b) Visualization results of Epicardium (Epi). The
 important regions marked in red show the most difference between prediction and ground
 truth.}
    \label{fig:sunny}
\end{figure}

Figure \ref{fig:sunny} shows representative results for segmentation of the cardiac MRI images from the Sunnybrook data. As can be observed in Fig. \ref{fig:sunny}a, our model produces better prediction results on the Endocardium set than the other models. For example, when predicting small sized objects, due to the combined use of MWL, our model produces predictions that match the ground truth while other models produce predictions that fall short. {Additionally, on the Epicardium dataset from Fig.\ref{fig:sunny}, our model achieves results that align much more closely with the ground truth compared to other models.}


For further assessments, Table \ref{tab:sunny} shows the quantitative evaluation metrics for both Endocardium and Epicardium regions from the left ventricle by the proposed LiteNeXt and other state-of-the-arts. {As can be seen from this table, our model achieves superior DSC and IoU results on both the Endocardium and Epicardium datasets while maintaining a significantly small number of parameters and GFLOPs, alongside relatively low inference latency and memory consumption.} The LiteNeXt gets the DSC of 94.05\%, IoU of 88.84\% for Endocardium; and DSC of 96.12\% , IoU of 92.60\% for Epicardium. {Meanwhile, the two lightweight models, UNeXt gets DSC of 92.71\% ($p$-value= 0.0001) and IoU 86.73\% ($p$-value= 0.0008) for Endocardium, and DSC of 94.83\% ($p$-value= 0.0385), IoU of 90.44\% ($p$-value= 0.0255) for Epicardial regions. Furthermore, our model also demonstrates superior performance across most metrics, with the $p$-values of the majority of comparison models being below 0.05, further validating the excellence of our model.}

\begin{table}[ht!]
\centering
\caption{The experiment comparison between different models on the Sunnybrook (SB) dataset.}
\label{tab:sunny}
\resizebox{\textwidth}{!}{%
\begin{tabular}{@{}lcccccccccccccc@{}}
\toprule
\multicolumn{14}{c}{\textbf{(a) Endocardium}} \\ \midrule
\multirow{2}{*}{\textbf{Models}} & 
\multirow{2}{*}{\textbf{Params}$\downarrow$} & 
\multirow{2}{*}{\textbf{FLOPS}$\downarrow$} & 
\multirow{2}{*}{\textbf{{gpuMem}}$\downarrow$} & 
\multirow{2}{*}{\textbf{{Infer}}$\downarrow$} & 
\multicolumn{2}{c}{\textbf{DSC (\%)$\uparrow$}} & 
\multicolumn{2}{c}{\textbf{IoU (\%)$\uparrow$}} & 
\multicolumn{2}{c}{\textbf{Precision (\%)$\uparrow$}} & 
\multicolumn{2}{c}{\textbf{Recall (\%)$\uparrow$}} & 
\textbf{F-score (\%)$\uparrow$} \\ 
\cmidrule(lr){6-7} \cmidrule(lr){8-9} \cmidrule(lr){10-11} \cmidrule(lr){12-13}
& & & & & Mean & {\textit{p}-value} & Mean & {\textit{p}-value} & Mean & {\textit{p}-value} & Mean & {\textit{p}-value}  \\ \midrule
UNet \cite{unet} & 31.13M & 55.84G & {655MiB} & {32.159ms} & {92.15} & {0.0002} & {87.75} & {0.0026} & {93.95} & {0.0876} & {92.49} & {0.0005} & {93.21} \\
UNet++ \cite{unet++} & 9.16M & 34.60G & {521MiB} & {27.167ms} & {93.18} & {0.0028} & {87.50} & {0.0029} & {91.87} & {0.0361} & {95.12} & {0.7195} & {93.47} \\
AttU-Net \cite{attentionunet} & 34.87M & 66.61G & {667MiB} & {38.239ms} & {91.73} & {0.0002} & {85.38} & {0.0005} & {90.75} & {0.0334} & {93.87} & {0.0027} & {92.28} \\
DCSAU-Net \cite{xu2023dcsau} & 2.60M & 6.91G & {491MiB} & {49.247ms} & {91.35} & {0.0010} & {84.70} & {3.190$\times10^{-5}$} & {94.41} & {0.1023} & {89.64} & {4.009$\times10^{-7}$} & {91.96} \\
SegNet \cite{badrinarayanan2017segnet} & 29.44M & 20.12G & {472MiB} & {27.631ms} & {90.21} & {1.281$\times10^{-4}$} & {82.64} & {9.154$\times10^{-6}$} & {93.03} & {0.8974} & {88.71} & {0.0002} & {90.82} \\
ResUnet \cite{resunet} & 13.04M & 68.07G & {675MiB} & {49.038ms} & {89.47} & {4.915$\times10^{-5}$} & {81.17} & {5.901$\times10^{-7}$} & {89.74} & {0.0127} & {90.78} & {3.533$\times10^{-7}$} & {90.25} \\
UNeXt \cite{valanarasu2022unext} & 1.47M & 0.52G & {331MiB} & {3.928ms} & {92.71} & {0.0001} & {86.73} & {0.0008} & {93.87} & {0.1485} & {92.30} & {0.0004} & {93.07} \\
nnUNet \cite{nnUnet} & {37.59M} & {0.44T} & {1015MiB} & {23768ms} & {92.44} & {0.0001} & {87.92} & {0.0022} & {\textbf{94.61}} & {0.0771} & {92.81} & {0.0007} & {93.70} \\
U-Netv2 \cite{unetv2} & {25.15M} & {5.58G} & {417MiB} & {57.425ms} & {93.65} & {0.5812} & {88.63} & {0.8421} & {94.02} & {0.2951} & {94.12} & {0.1901} & {94.07} \\
VM-UNetV2 \cite{vmUnetv2} & {22.77M} & {5.31G} & {451MiB} & {80.294ms} & {92.62} & {0.0087} & {86.77} & {0.0098} & {93.86} & {0.4001} & {92.38} & {0.0113} & {93.12} \\
LiteNeXt & 0.71M & 0.42G & {276MiB} & {1.892ms} & \textbf{94.05} & & \textbf{88.84} & & {93.10} & & \textbf{95.33} & & \textbf{94.18} \\ 
\midrule

\multicolumn{14}{c}{\textbf{(b) Epicardium}} \\ \midrule
\multirow{2}{*}{\textbf{Models}} & 
\multirow{2}{*}{\textbf{Params}$\downarrow$} & 
\multirow{2}{*}{\textbf{FLOPS}$\downarrow$} & 
\multirow{2}{*}{\textbf{{gpuMem}}$\downarrow$} & 
\multirow{2}{*}{\textbf{{Infer}}$\downarrow$} & 
\multicolumn{2}{c}{\textbf{DSC (\%)$\uparrow$}} & 
\multicolumn{2}{c}{\textbf{IoU (\%)$\uparrow$}} & 
\multicolumn{2}{c}{\textbf{Precision (\%)$\uparrow$}} & 
\multicolumn{2}{c}{\textbf{Recall (\%)$\uparrow$}} & 
\textbf{F-score (\%)$\uparrow$} \\ 
\cmidrule(lr){6-7} \cmidrule(lr){8-9} \cmidrule(lr){10-11} \cmidrule(lr){12-13}
& & & & & Mean & {\textit{p}-value} & Mean & {\textit{p}-value} & Mean & {\textit{p}-value} & Mean & {\textit{p}-value} \\ \midrule

UNet \cite{unet} & 31.13M & 55.84G & {655MiB} & {32.159ms} & {94.69} & {0.0164} & {90.24} & {0.0089} & {95.15} & {0.0877} & {94.43} & {0.0167} & {94.79} \\
UNet++ \cite{unet++} & 9.16M & 34.60G & {521MiB} & {27.167ms} & \textcolor{black}{95.06} & {0.0005} & \textcolor{black}{90.81} & {0.0003} & {95.92} & {0.2651} & {93.99} & {0.0091} & {94.95} \\
AttU-Net \cite{attentionunet} & 34.87M & 66.61G & {667MiB} & {38.239ms} & {95.09} & {0.0020} & {90.77} & {0.0013} & {96.18} & {0.5776} & {94.33} & {0.0107} & {95.25} \\
DCSAU-Net \cite{xu2023dcsau} & 2.60M & 6.91G & {491MiB} & {49.247ms} & {94.38} & {0.0218} & {89.77} & {0.0099} & {95.67} & {0.3413} & {93.36} & {0.0031} & {94.50} \\
SegNet \cite{badrinarayanan2017segnet} & 29.44M & 20.12G & {472MiB} & {27.631ms} & {90.91} & {2.421$\times10^{-6}$} & {83.66} & {5.769$\times10^{-7}$} & {94.41} & {0.0304} & {88.63} & {4.125$\times10^{-5}$} & {91.43} \\
ResUnet \cite{resunet} & 13.04M & 68.07G & {675MiB} & {49.038ms} & \textcolor{black}{93.45} & {0.0201} & \textcolor{black}{88.21} & {0.0217} & {96.61} & {0.9328} & {93.59} & {0.2299} & {94.33} \\
UNeXt \cite{valanarasu2022unext} & 1.47M & 0.52G & {331MiB} & {3.928ms} & {94.83} & {0.0385} & {90.44} & {0.0255} & {96.03} & {0.6167} & {93.94} & {0.0010} & {94.97} \\
{nnUNet} \cite{nnUnet} & {37.59M} & {0.44T} & {1015MiB} & {23768ms} & {94.67} & {0.0476} & {90.18} & {0.0492} & {93.68} & {0.0218} & {95.91} & {0.8917} & {93.78} \\
{U-Netv2} \cite{unetv2} & {25.15M} & {5.58G} & {417MiB} & {57.425ms} & {92.56} & {0.0156} & {90.51} & {0.0153} & {92.70} & {0.0116} & {92.96} & {0.2832} & {92.83} \\
{VM-UNetV2} \cite{vmUnetv2} & {22.77M} & {5.31G} & {451MiB} & {80.294ms} & {93.51} & {0.0201} & {90.56} & {0.0217} & {\textbf{96.61}} & {0.9328} & {93.59} & {0.2299} & {95.08} \\
LiteNeXt & 0.71M & 0.42G & {276MiB} & {1.892ms} & \textbf{96.12} & & \textbf{92.60} & & {96.50} & & \textbf{95.98} & & \textbf{96.24} \\ 
\bottomrule
\end{tabular}
}
\end{table}


\subsubsection{Evaluation on the Lung X-ray data}

To validate the generalizability of the model, we train the model on the Shenzhen dataset and test it on the Montgomery data. Besides, to validate the performance of the proposed LiteNeXt, we also conducted experiments with some other state-of-the-arts including: TransUNet \cite{transunet}, DoubleU-Net \cite{doubleunet} ,Unet \cite{unet},Unet++  and some lightweight models such as Ulite \cite{dinh20231m}, UNeXt \cite{valanarasu2022unext} , and MedT \cite{valanarasu2021medical}.

\begin{figure}[ht!]
    \centering
    \includegraphics[width=0.8\textwidth]{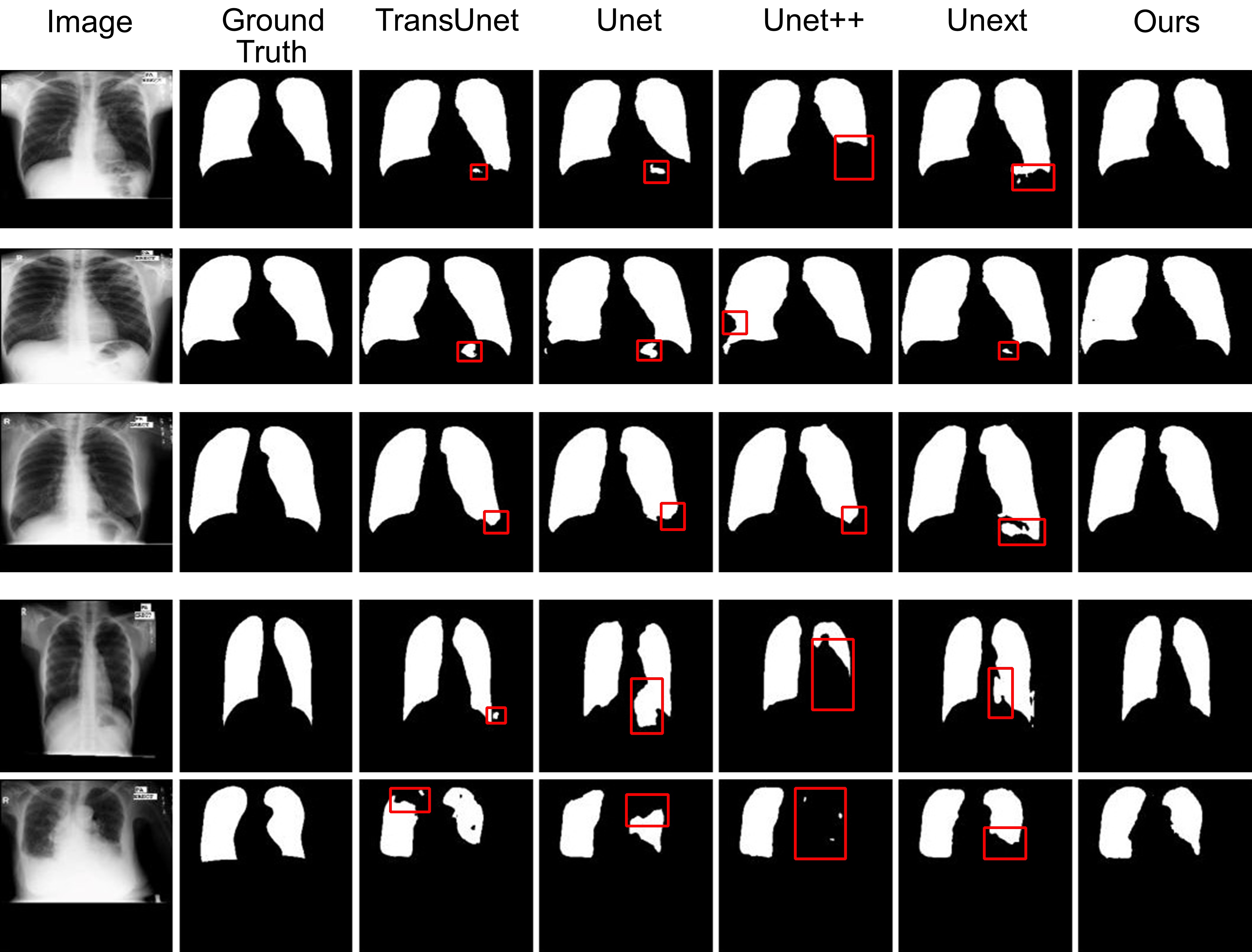}
    \caption{Representative visualization results of predictions on the Lung X-ray dataset. The
important regions marked in red show the most difference between prediction and ground
truth. }
    \label{fig:lung}
\end{figure}

Figure \ref{fig:lung} shows {representative} qualitative results for segmentation of the Lung  on the test set. {It is evident that our model delivers relatively outstanding results, with the two lungs segmented with high accuracy. Additionally, our model effectively minimizes the impact of noise and irrelevant regions in the images.}

\begin{table}[ht!]
\centering
\caption{{The experiment comparison between different models on the Lung X-ray dataset.}}
\label{tab:lung}
\resizebox{\textwidth}{!}{%
\begin{tabular}{@{}lccccccccccccccc@{}}
\toprule
\multirow{2}{*}{\textbf{Methods}} & 
\multirow{2}{*}{\textbf{Params}$\downarrow$} & 
\multirow{2}{*}{\textbf{FLOPS}$\downarrow$} & 
\multirow{2}{*}{\textbf{{gpuMem}}$\downarrow$} & 
\multirow{2}{*}{\textbf{{Infer}}$\downarrow$} & 
\multicolumn{2}{c}{\textbf{DSC (\%)$\uparrow$}} & 
\multicolumn{2}{c}{\textbf{IoU (\%)$\uparrow$}} & 
\multicolumn{2}{c}{\textbf{Precision (\%)$\uparrow$}} & 
\multicolumn{2}{c}{\textbf{Recall (\%)$\uparrow$}} & 
\multicolumn{1}{c}{\textbf{F-score (\%)$\uparrow$}} \\ 
\cmidrule(lr){6-7} \cmidrule(lr){8-9} \cmidrule(lr){10-11} \cmidrule(lr){12-13}
& & & & & Mean & {\textit{p}-value} & Mean & {\textit{p}-value} & Mean & {\textit{p}-value} & Mean & {\textit{p}-value}  \\ 
\midrule
TransUNet \cite{transunet} & \textcolor{black}{105.32M} & \textcolor{black}{38.52G} & {825MiB} & {44.746ms} & \textcolor{black}{96.03} & {0.0416} & \textcolor{black}{92.63} & {0.0356} & \textbf{\textcolor{black}{97.87}} & {0.2353} & \textcolor{black}{94.33} & {0.0249} & \textcolor{black}{96.07} \\
Double U-Net \cite{doubleunet} & \textcolor{black}{29.28M} & \textcolor{black}{53.81G} & {585MiB} & {49.247ms} & \textcolor{black}{96.03} & {0.0432} & \textcolor{black}{92.17} & {0.0332} & \textcolor{black}{96.63} & {0.0694} & \textcolor{black}{95.48} & {0.0431} & \textcolor{black}{96.05} \\
UNet \cite{unet} & \textcolor{black}{31.13M} & \textcolor{black}{55.84G} & {655MiB} & {32.159ms} & \textcolor{black}{96.09} & {0.0395} & \textcolor{black}{92.96} & {0.0399} & \textcolor{black}{96.27} & {0.0421} & \textcolor{black}{95.74} & {0.0412} & \textcolor{black}{96.00} \\
UNet++ \cite{unet++} & \textcolor{black}{9.16M} & \textcolor{black}{34.60G} & {521MiB} & {27.167ms} & \textcolor{black}{96.17} & {0.0262} & \textcolor{black}{92.85} & {0.0425} & \textcolor{black}{97.80} & {0.4218} & \textcolor{black}{94.47} & {0.0078} & \textcolor{black}{96.10} \\
UNeXt \cite{valanarasu2022unext} & \textcolor{black}{1.47M} & \textcolor{black}{0.52G} & {331MiB} & {3.928ms} & \textcolor{black}{96.19} & {0.0135} & \textcolor{black}{92.95} & {0.0199} & \textcolor{black}{96.59} & {0.0412} & \textcolor{black}{95.89} & {0.0072} & \textcolor{black}{96.24} \\
U-Lite \cite{dinh20231m} & \textcolor{black}{0.87M} & \textcolor{black}{0.54G} & {343MiB} & {4.708ms} & \textcolor{black}{95.61} & {0.0319} & \textcolor{black}{91.84} & {0.0014} & \textcolor{black}{96.16} & {0.0323} & \textcolor{black}{95.13} & {0.0049} & \textcolor{black}{95.64} \\
MedT \cite{valanarasu2021medical} & \textcolor{black}{1.60M} & \textcolor{black}{21.24G} & {423MiB} & {66.477ms} & \textcolor{black}{95.89} & {0.0067} & \textcolor{black}{92.01} & {0.0032} & \textcolor{black}{96.78} & {0.0483} & \textcolor{black}{95.21} & {0.0009} & \textcolor{black}{95.99} \\
{nnUNet} \cite{nnUnet} & {37.59M} & {0.44T} & {1015MiB} & {23768ms} & {95.04} & {0.0342} & {90.82} & {$7.948\times10^{-5}$} & {96.16} & {0.0099} & {94.21} & {$4.129\times10^{-5}$} & {95.18} \\
{U-Netv2} \cite{unetv2} & {25.15M} & {5.58G} & {417MiB} & {57.425ms} & {95.56} & {0.0407} & {91.74} & {0.0041} & {97.57} & {0.3929} & {93.86} & {$9.213\times10^{-6}$} & {95.68} \\
{VM-UNetV2} \cite{vmUnetv2} & {22.77M} & {5.31G} & {451MiB} & {80.294ms} & {95.29} & {0.0375} & {91.16} & {0.0769} & {92.28} & {$8.129\times10^{-6}$} & {96.10} & {0.1011} & {94.18} \\
\textbf{LiteNeXt} & \textbf{0.71M} & \textbf{0.42G} & \textbf{{276MiB}} & \textbf{{1.892ms}} & \textbf{96.70} & & \textbf{93.68} & & {97.22} & & \textbf{96.26} & & \textbf{96.74} \\ 
\bottomrule
\end{tabular}%
}
\end{table}


{The results in Table} \ref{tab:lung}  {showed that our LiteNeXt model outperformed other models on most metrics. Specifically, our method achieves mean DSC 96.70\% and mean IoU of 93.68\%, higher than those by other methods, especially the other lightweight models such as Ulite, UNeXt, MedT which achieve DSC scores of 95.61\% ($p$-value= 0.0319), 96.19\% ($p$-value= 0.0135), and 95.89\% ($p$-value= 0.0067), respectively. Additionally, the Recall, Precision, and F1-score results are 96.26\%, 97.22\%, and 96.74\%, respectively—remarkable outcomes given that our model's parameters and GFLOPs are exceptionally small, at only 0.71M and 0.42G. Moreover, the results consistently show $p$-values below 0.05, further validating the strength of our model.}

\subsection{Ablation Study}
\subsubsection{Performance of the proposed loss}

To find the weights $w_b, w_m, w_o$ for the proposed Marginal Weight Loss function, we perform imputation experiments on six datasets. With the assumption that $w_m > w_o > w_b$, $w_m + w_o + w_b = 1$ and the search step is $0.1$, four experimental cases were conducted and are presented in Table \ref{tab:ablation_weight}. Experimental results show that the weight set $w_b = 0.1, w_m=0.6, w_o=0.3$ gives the best average results on two metrics, Dice and IOU. Therefore this weight set can be used as the default weight set in the Marginal Weight Loss function when using it on different medical data sets.

\renewcommand{\arraystretch}{1.1}
\begin{table}[ht!]
\centering
\caption{Evaluation of using different weighting for Marginal Weight Loss. {DSC and IoU metrics are presented as percentages.}}
\label{tab:ablation_weight}
\resizebox{\textwidth}{!}{%
\begin{tabular}{lccccc ccccc ccccc cccc} 
\hline
\multirow{2}{*}{\textbf{\(w_b\)}} & 
\multirow{2}{*}{\textbf{\(w_m\)}} & 
\multirow{2}{*}{\textbf{\(w_o\)}} & 
\multicolumn{2}{c}{\textbf{Bowl2018}} & 
\multicolumn{2}{c}{\textbf{GlaS}} & 
\multicolumn{2}{c}{\textbf{ISIC2018}} & 
\multicolumn{2}{c}{\textbf{PH2}} & 
\multicolumn{2}{c}{\textbf{Endo}} & 
\multicolumn{2}{c}{\textbf{Epi}} & 
\multicolumn{2}{c}{\textbf{Lung}} \\ 
\cline{4-5} \cline{6-7} \cline{8-9} \cline{10-11} \cline{12-13} \cline{14-15} \cline{16-17}
& & & DSC$\uparrow$ & IoU$\uparrow$ & 
DSC$\uparrow$ & IoU$\uparrow$ & 
DSC$\uparrow$ & IoU$\uparrow$ & 
DSC$\uparrow$ & IoU$\uparrow$ & 
DSC$\uparrow$ & IoU$\uparrow$ & 
DSC$\uparrow$ & IoU$\uparrow$ & 
DSC$\uparrow$ & IoU$\uparrow$ \\ \hline
0.1 & 0.5 & 0.4 & 92.35 & 86.23 & 90.38 & 83.92 & 90.22 & 83.64 & 94.87 & 90.90 & 93.33 & 88.82 & 95.56 & 92.12 & 96.02 & 92.32 \\
\textbf{0.1} & \textbf{0.6} & \textbf{0.3} & \textbf{92.50} & \textbf{86.39} & \textbf{90.91} & \textbf{84.15} & \textbf{90.52} & \textbf{83.93} & \textbf{95.18} & \textbf{90.98} & \textbf{94.05} & 88.84 & 96.14 & \textbf{92.60} & \textbf{96.70} & \textbf{93.68} \\
0.1 & 0.7 & 0.2 & 92.30 & 86.11 & 90.25 & 83.67 & 90.27 & 83.55 & 94.63 & 89.52 & 93.38 & \textbf{88.87} & \textbf{96.17} & 92.51 & 95.95 & 92.12 \\ 
0.2 & 0.5 & 0.3 & 92.19 & 86.02 & 90.09 & 83.01 & 90.32 & 83.94 & 94.54 & 89.30 & 93.36 & 88.75 & 95.49 & 91.83 & 96.11 & 92.78 \\ \hline
\end{tabular}
}
\end{table}



{To select the size of the kernel size in MWL Loss, we conducted experiments on 7 kernel sizes: 1, 3, 5, 7, 9, 11, 13. With a kernel size of 1, it is equivalent to using the B-BCE Loss function. The experimental results from Table} \ref{tab:ablation_kernel_size} {show that the segmentation accuracy tends to increase as the kernel size increases and reaches the highest accuracy with a kernel size of 9 on all test sets. However, as the kernel size continues to increase, the segmentation accuracy tends to decrease. This is because the kernel sizes that are too large have significantly narrowed the weight regions of the objects on the weight mask, causing them to be unbalanced during the learning process.}

\renewcommand{\arraystretch}{1.1}
\begin{table}[ht!]
\centering
\caption{Evaluation of using different kernel sizes for Marginal Weight Loss. {DSC and IoU metrics are presented as percentages.}}
\label{tab:ablation_kernel_size}
\resizebox{\textwidth}{!}{%
\begin{tabular}{l ccccc ccccc ccccc cc} 
\hline
\multirow{2}{*}{\textbf{k}} & 
\multicolumn{2}{c}{\textbf{Bowl2018}} & 
\multicolumn{2}{c}{\textbf{GlaS}} & 
\multicolumn{2}{c}{\textbf{ISIC2018}} & 
\multicolumn{2}{c}{\textbf{PH2}} & 
\multicolumn{2}{c}{\textbf{Endo}} & 
\multicolumn{2}{c}{\textbf{Epi}} & 
\multicolumn{2}{c}{\textbf{Lung}} \\ 
\cline{2-3} \cline{4-5} \cline{6-7} \cline{8-9} \cline{10-11} \cline{12-13} \cline{14-15}
& DSC$\uparrow$ & IoU$\uparrow$ & 
DSC$\uparrow$ & IoU$\uparrow$ & 
DSC$\uparrow$ & IoU$\uparrow$ & 
DSC$\uparrow$ & IoU$\uparrow$ & 
DSC$\uparrow$ & IoU$\uparrow$ & 
DSC$\uparrow$ & IoU$\uparrow$ & 
DSC$\uparrow$ & IoU$\uparrow$ \\ \hline
1 & 91.93 & 85.31 & 90.20 & 82.96 & 90.08 & 83.52 & 94.38 & 89.66 & 93.29 & 87.75 & 95.24 & 91.22 & 96.26 & 92.93 \\
3 & 91.99 & 85.42 & 90.22 & 83.11 & 89.97 & 83.48 & 94.57 & 89.86 & 93.38 & 87.83 & 95.77 & 91.89 & 96.31 & 93.08 \\
5 & 91.92 & 85.27 & 90.38 & 83.87 & 90.12 & 83.92 & 94.58 & 89.72 & 93.59 & 87.66 & 95.62 & 91.88 & 96.22 & 93.07 \\
7 & 92.17 & 85.98 & 90.74 & 84.02 & 90.33 & \textbf{83.97} & 94.77 & 90.33 & 94.02 & \textbf{88.89} & 95.95 & 92.17 & 96.48 & 93.22 \\
\textbf{9} & \textbf{92.50} & \textbf{86.39} & \textbf{90.91} & \textbf{84.15} & \textbf{90.52} & 83.93 & \textbf{95.18} & \textbf{90.98} & \textbf{94.05} & 88.87 & \textbf{96.17} & \textbf{92.60} & \textbf{96.70} & \textbf{93.68} \\
11 & 92.11 & 85.79 & 90.77 & 84.11 & 90.43 & 83.87 & 94.82 & 90.78 & 94.00 & 88.41 & 96.01 & 92.38 & 96.61 & 93.61 \\
13 & 91.89 & 85.11 & 90.03 & 82.64 & 89.34 & 82.97 & 94.33 & 89.21 & 93.11 & 87.25 & 95.26 & 91.28 & 95.89 & 92.07 \\ \hline
\end{tabular}
}
\end{table}

\renewcommand{\arraystretch}{0.9}
\begin{table}[ht!]
\centering
\caption{Comparison between different loss functions on the six datasets: Bowl2018, GLaS, ISIC2018, PH2, Sunnybrook (SB), and Lung. {Metrics are presented as percentages.}}
\label{tab:ablation_loss}
\begin{tabular}{llccccc}
\hline
\textbf{Dataset} & \textbf{Loss} & \textbf{DSC$\uparrow$} & \textbf{IoU$\uparrow$} & \textbf{Precision$\uparrow$} & \textbf{Recall$\uparrow$} & \textbf{F-score$\uparrow$} \\ \hline
\multirow{7}{*}{Bowl2018} & $L_{Focal loss}$ & 91.45 & 84.72 & 91.58 & 91.59 & 91.58 \\
 & $L_{BCE}$ & 91.68 & 85.07 & 91.92 & 91.92 & 91.92 \\
 & $L_{W-BCE}$ & 91.77 & 85.23 & 91.69 & 92.82 & 92.25 \\
 & $L_{B-BCE}$ & 91.98 & 85.38 & 91.78 & 92.51 & 92.14 \\
 & $L_{MWL}$ & 92.17 & 85.85 & 91.31 & 93.49 & 92.38 \\
 & $L_{BCE-SeRP-Dice}$ & 92.35 & 86.13 & \textbf{92.36} & 92.75 & 92.55 \\
 & $L_{MWL-SeRP-Dice}$ & \textbf{92.50} & \textbf{86.39} & 91.85 & \textbf{93.50} & \textbf{92.70} \\ \cline{2-7}
\multirow{7}{*}{GLaS} & $L_{Focal loss}$ & 89.68 & 82.09 & 90.94 & 89.42 & 90.17 \\
 & $L_{BCE}$ & 89.95 & 82.53 & 89.67 & 91.19 & 90.42 \\
 & $L_{W-BCE}$ & 89.08 & 81.17 & 90.21 & 89.61 & 89.91 \\
 & $L_{B-BCE}$ & 89.94 & 82.46 & 90.27 & 89.91 & 90.09 \\
 & $L_{MWL}$ & 90.09 & 82.77 & 90.19 & 89.98 & 90.08 \\
 & $L_{BCE-SeRP-Dice}$ & 90.17 & 82.88 & 88.88 & \textbf{92.58} & 90.69 \\
 & $L_{MWL-SeRP-Dice}$ & \textbf{90.91} & \textbf{84.15} & \textbf{91.73} & 90.89 & \textbf{91.31} \\ \cline{2-7}
\multirow{7}{*}{ISIC2018} & $L_{Focal loss}$ & 89.69 & 82.65 & 90.35 & 91.07 & 90.71 \\
 & $L_{BCE}$ & 89.57 & 83.29 & 90.77 & 88.97 & 89.86 \\
 & $L_{W-BCE}$ & 89.97 & 83.43 & 89.87 & 90.28 & 90.07 \\
 & $L_{B-BCE}$ & 89.89 & 83.39 & 90.18 & 91.21 & 90.69 \\
 & $L_{MWL}$ & 90.12 & 83.87 & 91.13 & 90.98 & 91.05 \\
 & $L_{BCE-SeRP-Dice}$ & 90.15 & 83.62 & \textbf{93.10} & 89.98 & 91.51 \\
 & $L_{MWL-SeRP-Dice}$ & \textbf{90.52} & \textbf{83.93} & 92.00 & \textbf{91.29} & \textbf{91.64} \\ \cline{2-7}
\multirow{7}{*}{PH2} & $L_{Focal loss}$ & 92.73 & 86.66 & \textbf{95.72} & 90.63 & 93.10 \\
 & $L_{BCE}$ & 94.28 & 89.32 & 93.91 & 95.09 & 94.49 \\
 & $L_{W-BCE}$ & 94.31 & 89.52 & 94.44 & 95.12 & 94.78 \\
 & $L_{B-BCE}$ & 94.29 & 89.23 & 93.89 & \textbf{95.41} & 94.64 \\
 & $L_{MWL}$ & 94.37 & 89.78 & 94.35 & 94.97 & 94.66 \\
 & $L_{BCE-SeRP-Dice}$ & 94.44 & 89.66 & 94.29 & 95.12 & 94.70 \\
 & $L_{MWL-SeRP-Dice}$ & \textbf{95.18} & \textbf{90.98} & 95.42 & 95.33 & \textbf{95.37} \\ \cline{2-7}
\multirow{7}{*}{SB Endo} & $L_{Focal loss}$ & 93.23 & 87.62 & 93.64 & 93.37 & 93.50 \\
 & $L_{BCE}$ & 92.93 & 87.25 & 92.72 & 93.75 & 93.23 \\
 & $L_{W-BCE}$ & 93.49 & 88.22 & \textbf{94.77} & 92.89 & 93.82 \\
 & $L_{B-BCE}$ & 93.25 & 87.58 & 94.45 & 92.86 & 93.64 \\
 & $L_{MWL}$ & 93.27 & 87.60 & 93.59 & 93.59 & 93.59 \\
 & $L_{BCE-SeRP-Dice}$ & 93.40 & 87.85 & 93.94 & 93.53 & 93.73 \\
 & $L_{MWL-SeRP-Dice}$ & \textbf{94.06} & \textbf{88.84} & 93.11 & \textbf{95.33} & \textbf{94.21} \\ \cline{2-7}
\multirow{7}{*}{SB Epi} & $L_{Focal loss}$ & 95.30 & 91.24 & 95.18 & 95.64 & 95.41 \\
 & $L_{BCE}$ & 95.37 & 91.22 & 96.46 & 94.53 & 95.48 \\
 & $L_{W-BCE}$ & 95.24 & 91.12 & 95.73 & 95.01 & 95.37 \\
 & $L_{B-BCE}$ & 95.34 & 91.21 & 96.05 & 94.37 & 95.20 \\
 & $L_{MWL}$ & 95.46 & 91.51 & 96.11 & 95.03 & 95.56 \\
 & $L_{BCE-SeRP-Dice}$ & 95.49 & 91.49 & 96.37 & 94.74 & 95.54 \\
 & $L_{MWL-SeRP-Dice}$ & \textbf{96.14} & \textbf{92.60} & \textbf{96.51} & \textbf{95.89} & \textbf{96.20} \\ \cline{2-7}
\multirow{7}{*}{Lung} & $L_{Focal loss}$ & 95.54 & 91.59 & 95.54 & 95.70 & 95.62 \\
 & $L_{BCE}$ & 95.40 & 91.32 & 94.96 & 96.00 & 95.47 \\
 & $L_{W-BCE}$ & 96.03 & 92.48 & 94.79 & 97.43 & 96.09 \\
 & $L_{B-BCE}$ & 95.99 & 92.42 & 95.92 & 96.22 & 96.07 \\
 & $L_{MWL}$ & 96.27 & 92.95 & 96.43 & 96.26 & 96.34 \\
 & $L_{BCE-SeRP-Dice}$ & 96.40 & 93.17 & 96.33 & 96.62 & 96.47 \\
 & $L_{MWL-SeRP-Dice}$ & \textbf{96.70} & \textbf{93.68} & \textbf{97.22} & \textbf{96.26} & \textbf{96.74} \\ \hline
\end{tabular}
\end{table}

To assess the performance of the proposed loss, we compare the  different loss functions used in training our proposed model for the {six} datasets in Table \ref{tab:ablation_loss}. The comparative losses include: the Focal loss ($L_{Focalloss}$)\cite{focalloss}, Binary Cross Entropy ($L_{BCE}$)\cite{BCE}, Weighted Binary Cross Entropy ($L_{W-BCE}$)\cite{WBCE}, Balanced Binary Cross Entropy ($L_{B-BCE}$)\cite{BBCE}, our proposed Marginal Weight Loss ($L_{MWL}$), Combination of BCE with  SeRP and Dice ($L_{BCE-SeRP-Dice}$), and the combination of MWL and SeRP with Dice ($L_{MWL-SePR-Dice}$). As can be observed in Table \ref{tab:ablation_loss}, our proposed MWL slightly outperforms other weighted loss functions such as W-BCE and B-BCE on all datasets. For example, DCS for Bowl2018 is 92.17\% when training with MWL while it is 91.77\% when training with W-BCE. We also demonstrate the efficiency of SeRP when combining MWL with SeRP and Dice loss. The results in Table \ref{tab:ablation_loss} show that this combination provides superior performance compared to all other losses when performed on all datasets. For example, the DSC for ISIC2018 is 90.52\% while trained on the proposed loss, while it is 89.57\% by using BCE loss. The IoU by the proposed loss is 84.15\% while the BCE obtains the IoU of 82.53\% for the GLaS data. {With the Lung dataset, the proposed loss function gives a DSC of 96.70\% while the DSC is only 95.40\% when using BCE Loss.}.

\subsubsection{Performance of the proposed LGEMixer }

To evaluate the performance of the proposed LGEMixer module for feature extraction, we conduct the experiments to compare with those when using the previous modules including ConvMixer and ConvNeXt. In particular, in the proposed model, we only replace the LGE blocks in Fig. \ref{LiteNeXt}a and  Fig. \ref{LiteNeXt}b by the ConvMixer and ConvNeXt blocks in the corresponding experiments. The same training pipeline is applied for experiments with the ConvMixer modules, ConvNeXt modules and LGEMixer modules { in Fig. \ref{block_lge_compare}}.

\renewcommand{\arraystretch}{1.1}
\begin{table}[ht!]
\centering
\caption{Performance of the proposed model when using different feature extraction modules: ConvMixer, ConvNeXt, and LGEMixer, when evaluated on six datasets: Bowl2018, GlaS, ISIC2018, PH2, Sunnybrook (SB), and Lung. {DSC and IoU metrics are presented as percentages.}}
\label{tab:ablation_feature}
\begin{tabular}{lccccc}
\hline
\textbf{Dataset} & \textbf{Experiment Ablation} & \textbf{Params}$\downarrow$ & \textbf{FLOPS}$\downarrow$ & \textbf{DSC}$\uparrow$ & \textbf{IoU}$\uparrow$ \\ \hline
\multirow{3}{*}{Bowl2018} & ConvMixer & 2.03M & 1.04G & 91.17 & 84.17 \\
 & ConvNeXt & 2.63M & 0.94G & 90.31 & 82.74 \\
 & LGEMixer (Ours) & 0.71M & 0.42G & \textbf{92.50} & \textbf{86.39} \\ \cline{2-6}
\multirow{3}{*}{GlaS} & ConvMixer & 2.03M & 1.04G & 88.08 & 79.54 \\
 & ConvNeXt & 2.63M & 0.94G & 88.29 & 80.00 \\
 & LGEMixer (Ours) & 0.71M & 0.42G & \textbf{90.91} & \textbf{84.15} \\ \cline{2-6}
\multirow{3}{*}{ISIC2018} & ConvMixer & 2.03M & 1.04G & 90.25 & 83.50 \\
 & ConvNeXt & 2.63M & 0.94G & 89.91 & 83.17 \\
 & LGEMixer (Ours) & 0.71M & 0.42G & \textbf{90.52} & \textbf{83.93} \\ \cline{2-6}
\multirow{3}{*}{PH2} & ConvMixer & 2.03M & 1.04G & 94.04 & 88.97 \\
 & ConvNeXt & 2.63M & 0.94G & 94.39 & 89.58 \\
 & LGEMixer (Ours) & 0.71M & 0.42G & \textbf{95.18} & \textbf{90.98} \\ \cline{2-6}
\multirow{3}{*}{SB Endo} & ConvMixer & 2.03M & 1.04G & 90.60 & 83.61 \\
 & ConvNeXt & 2.63M & 0.94G & 92.99 & 87.19 \\
 & LGEMixer (Ours) & 0.71M & 0.42G & \textbf{94.06} & \textbf{88.84} \\ \cline{2-6}
\multirow{3}{*}{SB Epi} & ConvMixer & 2.03M & 1.04G & 90.25 & 84.86 \\
 & ConvNeXt & 2.63M & 0.94G & 94.91 & 90.43 \\
 & LGEMixer (Ours) & 0.71M & 0.42G & \textbf{96.14} & \textbf{92.60} \\ \cline{2-6}
\multirow{3}{*}{Lung} & ConvMixer & 2.03M & 1.04G & 95.21 & 91.78 \\
 & ConvNeXt & 2.63M & 0.94G & 95.99 & 92.35 \\
 & LGEMixer (Ours) & 0.71M & 0.42G & \textbf{96.70} & \textbf{93.68} \\ \hline
\end{tabular}
\end{table}

As can be seen from Table \ref{tab:ablation_feature}, for the Bowl2018 data, the DSC when using LGEMixer is 92.5\% while the score is 90.31\% when using ConvNeXt. One of the most significant differences is the IoU values in GlaS dataset. The model when using the LGEMixer obtains the IoU value of 84.15\% while it gets 79.54\%, and 80\% when using ConvMixer, ConvNeXt respectively. Another competitive result is that on the Lung dataset, the model with LGEMixer block gives 1.9\% and 1.33\% better IoU than when using ConvMixer block and ConvNeXt respectively.  It is worth mentioning that using the LGEMixer also helps reduce the number of parameters. The proposed framework only has 0.71M parameters, while it will be 2.03M and 2.63M if using respectively ConvMixer and ConvNeXt. 

\subsubsection{Performance of the Self-embedding Representation Parallel}


To show the performance of our Self-embedding Representation Parallel (SeRP) algorithm, we compare the results for all datasets when using (w SeRP) and without using (w/o SeRP) in Table \ref{tab:ablation_SeRP}. {The comparison demonstrates significant improvements across multiple metrics when employing SeRP.} For example, with the Endocardium data, the average DSC score is 94.06\% when using SeRP, compared with 93.36\% in the case of without using SeRP. For GlaS data, the mean IoU without using SeRP is 82.98\%, much lower than that when using SeRP, which achieves 84.15\%.

\renewcommand{\arraystretch}{1.1} 
\begin{table}[ht!]
\centering
\caption{Performance of our Self-embedding Representation Parallel (SeRP) algorithm. We compare the results for the six datasets when using (w SeRP) and without using (w/o SeRP). {Metrics are presented as percentages.}}
\label{tab:ablation_SeRP}
\begin{tabular}{lcccccc}
\hline
\textbf{Dataset} & \textbf{Experiment Ablation} & \textbf{DSC}$\uparrow$ & \textbf{IoU}$\uparrow$ & \textbf{Precision}$\uparrow$ & \textbf{Recall}$\uparrow$ & \textbf{F-score}$\uparrow$ \\ \hline
\multirow{2}{*}{Bowl2018} & w SeRP & \textbf{92.50} & \textbf{86.39} & \textbf{91.85} & \textbf{93.50} & \textbf{92.70} \\
 & w/o SeRP & 92.22 & 85.97 & 91.66 & 92.84 & 92.24 \\ \cline{2-7}
\multirow{2}{*}{GlaS} & w SeRP & \textbf{90.91} & \textbf{84.15} & \textbf{91.73} & \textbf{90.89} & \textbf{91.31} \\
 & w/o SeRP & 90.12 & 82.98 & 89.59 & 90.63 & 90.11 \\ \cline{2-7}
\multirow{2}{*}{ISIC2018} & w SeRP & \textbf{90.52} & \textbf{83.93} & \textbf{92.00} & \textbf{91.29} & \textbf{91.64} \\
 & w/o SeRP & 90.29 & 83.72 & 93.35 & 89.86 & 91.57 \\ \cline{2-7}
\multirow{2}{*}{PH2} & w SeRP & \textbf{95.18} & \textbf{90.98} & \textbf{95.42} & \textbf{95.33} & \textbf{95.37} \\
 & w/o SeRP & 94.39 & 89.27 & 95.40 & 94.25 & 94.82 \\ \cline{2-7}
\multirow{2}{*}{SB Endo} & w SeRP & \textbf{94.06} & \textbf{88.84} & \textbf{93.11} & \textbf{95.33} & \textbf{94.21} \\
 & w/o SeRP & 93.36 & 88.22 & 92.16 & 95.01 & 93.56 \\ \cline{2-7}
\multirow{2}{*}{SB Epi} & w SeRP & \textbf{96.14} & \textbf{92.60} & \textbf{96.51} & \textbf{95.89} & \textbf{96.20} \\
 & w/o SeRP & 95.47 & 91.79 & 96.22 & 95.23 & 95.56 \\ \cline{2-7}
\multirow{2}{*}{{Lung}} & {w SeRP} & \textbf{{96.70}} & \textbf{{93.68}} & \textbf{{97.22}} & \textbf{{96.26}} & \textbf{{96.74}} \\
 & {w/o SeRP} & {96.48} & {93.27} & {96.82} & {96.20} & {96.51} \\ \hline
\end{tabular}
\end{table}

\renewcommand{\arraystretch}{1.1}
\begin{table}[ht!]
\centering
\caption{Evaluation of using different distances for the Self-embedding Representation Parallel (SeRP) algorithm. {DSC and IoU metrics are presented as percentages.}}
\label{tab:ablation_distances}
\begin{tabular}{lccccc}
\hline
\textbf{Dataset} & \textbf{Result} & \multicolumn{1}{c}{\textbf{Euclid}} & \multicolumn{1}{c}{\textbf{Manhattan}} & \multicolumn{1}{c}{\textbf{KL}} & \multicolumn{1}{c}{\textbf{Cosine}} \\ 
 &  & \textbf{Distance} & \textbf{Distance} & \textbf{Distance} & \textbf{Similarity} \\ \hline
\multirow{2}{*}{Bowl2018} & DSC$\uparrow$ & 92.34 & 92.38 & 92.42 & \textbf{92.50} \\
 & IoU$\uparrow$ & 86.12 & 86.65 & 86.25 & \textbf{86.39} \\ \cline{2-6}
\multirow{2}{*}{GlaS} & DSC$\uparrow$ & 90.25 & 90.47 & 89.63 & \textbf{90.91} \\
 & IoU$\uparrow$ & 82.89 & 83.20 & 82.08 & \textbf{84.15} \\ \cline{2-6}
\multirow{2}{*}{ISIC2018} & DSC$\uparrow$ & 90.36 & 90.31 & 90.19 & \textbf{90.52} \\
 & IoU$\uparrow$ & 83.79 & 83.75 & 83.46 & \textbf{83.93} \\ \cline{2-6}
\multirow{2}{*}{PH2} & DSC$\uparrow$ & 94.84 & 94.67 & 94.94 & \textbf{95.18} \\
 & IoU$\uparrow$ & 90.37 & 90.05 & 90.54 & \textbf{90.98} \\ \cline{2-6}
\multirow{2}{*}{SB Endo} & DSC$\uparrow$ & 93.17 & 93.52 & 93.56 & \textbf{94.06} \\
 & IoU$\uparrow$ & 87.43 & 87.94 & 88.06 & \textbf{88.84} \\ \cline{2-6}
\multirow{2}{*}{SB Epi} & DSC$\uparrow$ & 95.52 & 95.70 & 95.81 & \textbf{96.14} \\
 & IoU$\uparrow$ & 91.70 & 91.85 & 92.01 & \textbf{92.60} \\ \cline{2-6}
\multirow{2}{*}{Lung} & DSC$\uparrow$ & 96.54 & 96.42 & 96.44 & \textbf{96.70} \\
 & IoU$\uparrow$ & 93.41 & 93.17 & 93.23 & \textbf{93.68} \\ \hline
\end{tabular}
\end{table}

{To further evaluate the impact of different distance measures used in the SeRP algorithm, experiments were conducted comparing Euclid, Manhattan, Kullback-Leibler (KL), and Cosine similarity distances. Table \ref{tab:ablation_distances} shows the results across the datasets.} As can be seen in the last column of Table \ref{tab:ablation_distances}, using the Cosine similarity give the best scores for both DSC and IoU with the six datasets. For example, the Cosine similarity (CS) gets IoU of 84.15\% for GlaS data, while the common Euclid distance gets the value of 82.89\% for this dataset. For the PH2 data, the CS obtains the DSC of 95.18\% while the KL distance gives 94.94\%, Manhattan distances gives the score of 94.67\%, and Euclid distance obtains the DSC of 94.84\%. Notably, for the Epicardium of Sunnybrook, the IoU by Cosine similarity is 92.60\%, while the values are 91.7\% with Euclid distance, and 91.85\% when using the Manhattan distance. For the Lung dataset, as the last row in Table \ref{tab:ablation_distances}, the Cosine similarity for IoU is 93.68\% while the Manhattan distance and KL distance are only 93.17\% and 93.23\% respectively.


\section{Conclusion}
In this study, we introduce a lightweight model with a novel design for segmentation of medical images. We introduced a new module for feature extraction namely LGEMixer, so that the model encoder can get better information from the input image, while the decoder is simplified. We proposed a new loss, Marginal weight loss that has shown excellent performance for microscopic images. Furthermore, inspired from the self supervised learning, we introduce a new technique called Self-embedding Representation Parallel (SeRP) for the image segmentation task. The SeRP training strategy is also evaluated with different distances like Euclid, Manhattan, Kullback Leibler, and Cosine similarity distances. Experiments on various image modalities show that  though having very small parameter numbers and GFLOPs, our approach has comparative performances compared with other state of the arts. In the future work, acknowledging the significant role of annotated data in medical imaging, which often necessitates substantial resources, including the expertise of experienced clinicians, we recognize the challenges posed by the reliance on labeled data for deep learning models. Self-supervised learning provides a method to extract more information from limited labeled data. Looking forward, the application of semi-supervised learning emerges as a promising direction, as it mitigates the dependence on extensive labeling. This approach not only reduces the cost associated with data annotation but also lessens the reliance on labeled data, potentially enhancing the efficiency and applicability of deep learning models in medical imaging.

\section*{Acknowledgement}
This research is funded by Vietnam National Foundation for Science and Technology Development (NAFOSTED) under grant number 102.05-2021.34.



\end{document}